\documentclass[prb,twocolumn,aps,superscriptaddress,showpacs]{revtex4-1}

\usepackage{amsmath,amssymb}
\usepackage{graphicx}
\usepackage{xcolor}
\usepackage{graphicx}
\DeclareGraphicsExtensions{.pdf,.eps,.png,.jpg} \graphicspath{{./figs/}}
\usepackage{dcolumn}
\usepackage{bm}
\usepackage{tabularx}

\begin{document}

\title{
Spin-glass state and reversed magnetic anisotropy induced by Cr doping \\
in the Kitaev magnet $\alpha$-RuCl$_3$}

\author{G. Bastien}
\email{g.bastien@ifw-dresden.de} \affiliation{Leibniz-Institut f\"ur Festk\"orper- und
Werkstoffforschung (IFW) Dresden, 01171 Dresden, Germany}
\author{M. Roslova}
\affiliation{Fakult\"at f\"ur Chemie und Lebensmittelchemie, Technische Universit\"at Dresden,
01062 Dresden, Germany}
\affiliation{Department of Materials and Environmental Chemistry, Stockholm University, Stockholm SE-10691, Sweden}
\author{M. H. Haghighi}
\affiliation{Leibniz-Institut f\"ur Festk\"orper- und Werkstoffforschung (IFW) Dresden, 01171
Dresden, Germany}
\author{K. Mehlawat}
\affiliation{Leibniz-Institut f\"ur Festk\"orper- und Werkstoffforschung (IFW) Dresden, 01171
Dresden, Germany}
\author{J. Hunger}
\affiliation{Fakult\"at f\"ur Chemie und Lebensmittelchemie, Technische Universit\"at Dresden,
01062 Dresden, Germany}
\author{A. Isaeva}
\affiliation{Leibniz-Institut f\"ur Festk\"orper- und Werkstoffforschung (IFW) Dresden, 01171
Dresden, Germany}
\affiliation{Institut f\"ur Festk\"orper- und Materialphysik, Technische
Universit\"at Dresden, 01062 Dresden, Germany}
\author{T. Doert}
\affiliation{Fakult\"at f\"ur Chemie und Lebensmittelchemie, Technische Universit\"at Dresden,
01062 Dresden, Germany}
\author{M. Vojta}
\affiliation{Institut f\"ur Theoretische Physik, Technische Universit\"at Dresden, 01062 Dresden,
Germany}
\author{B. B\"uchner}
\affiliation{Leibniz-Institut f\"ur Festk\"orper- und Werkstoffforschung (IFW) Dresden, 01171
Dresden, Germany}
\affiliation{Institut f\"ur Festk\"orper- und Materialphysik, Technische
Universit\"at Dresden, 01062 Dresden, Germany}
\author{A. U. B. Wolter}
\affiliation{Leibniz-Institut f\"ur Festk\"orper- und Werkstoffforschung (IFW) Dresden, 01171
Dresden, Germany}

\date{\today}

\begin{abstract}

Magnetic properties of the substitution series Ru$_{1-x}$Cr$_x$Cl$_3$ were investigated to determine the evolution from the anisotropic Kitaev magnet $\alpha$-RuCl$_3$ with $J_{\rm eff} = 1/2$ magnetic Ru$^{3+}$ ions to the isotropic Heisenberg magnet CrCl$_3$ with $S = 3/2$ magnetic
Cr$^{3+}$ ions.
Magnetization measurements on single crystals revealed a reversal of the magnetic
anisotropy under doping, which we argue to arise from the competition between anisotropic Kitaev and off-diagonal interactions on the Ru-Ru links and approximately isotropic Cr-Ru and isotropic Cr-Cr interactions.
In addition, combined magnetization, ac susceptibility and specific-heat
measurements clearly show the destabilization of the long-range magnetic order of $\alpha$-RuCl$_3$ in favor of a spin-glass state of Ru$_{1-x}$Cr$_x$Cl$_3$ for a
low doping of $x\backsimeq0.1$. The corresponding freezing temperature as a function of Cr content shows a broad maximum around $x\backsimeq0.45$.

\end{abstract}


\maketitle

\section{Introduction}

The Kitaev model on a honeycomb lattice realizes a topological spin liquid in two space dimensions with fractionalized Majorana-fermion excitations~\cite{Kitaev2006,Bonderson2008, Knolle2015}. The model is constructed from bond-dependent Ising interactions $K$ between nearest-neighbor moments which induce strong exchange frustration. Crucial ingredients for realizing the Kitaev model in materials are strong spin-orbit coupling combined with cancellation tendencies between multiple exchange paths~\cite{Jackeli2009}.
The presence of dominant Kitaev interactions was first suggested for
honeycomb iridates~\cite{cao2013} and later for $\alpha$-RuCl$_3$, based on an unusual magnetic
excitation spectrum~\cite{Banerjee2016, Banerjee2017}, strong magnetic
anisotropy~\cite{Majumder2015}, and electronic-structure calculations~\cite{Winter2016,Yadav2016}.

$\alpha$-RuCl$_3$ is a $J_{\rm eff}=1/2$ Mott insulator with a layered structure of edge-sharing
(RuCl$_6$) octahedra forming a honeycomb lattice~\cite{Fletcher1967}. These van-der-Waals-bonded
honeycomb layers are arranged in a monoclinic unit cell at room temperature (space group $C2/m$)~\cite{Johnson2015},
however, the stacking of these layers may be inhomogeneous depending on the crystal
quality~\cite{Cao2016}. Apart from a dominant Kitaev exchange interaction, additional magnetic
interactions such as nearest and next-nearest neighbor Heisenberg interactions $J$ and symmetric off-diagonal couplings $\Gamma$ have to be considered in a minimal model to describe the physics of $\alpha$-RuCl$_3$~\cite{Winter2016,Yadav2016,Janssen2017,Winter2018}.
Due to these additional terms in the Hamiltonian, $\alpha$-RuCl$_3$ orders in an antiferromagnetic zigzag state at low temperature~\cite{Johnson2015, Cao2016}. Two magnetic transitions were identified at $T_{N}\backsimeq$ 7~K and $T_{N}\backsimeq$ 13~K in Ref.~\onlinecite{Cao2016}, based on a $C2/m$ space group, with magnetic wave vectors (0, 1, 1/3) (ABC stacking) and (0, 1,1/2) (AB stacking), respectively. These orders can coexist in the same single crystal, which may contain several magnetic domains ~\cite{Majumder2015, Kubota2015, Cao2016}.
Although $\alpha$-RuCl$_3$ has an antiferromagnetic ground state, it is considered as a proximate Kitaev system since signatures of Majorana
fermions were recently reported in the paramagnetic state above $T_N$~\cite{Banerjee2016, Do2017, Jansa2018, Wellm2018}.
Furthermore, magnetic fields of about $\mu_0H_c\backsimeq$ 7-8~T suppress long-range magnetic order toward a quantum paramagnetic state which has been proposed to realize a quantum spin liquid~\cite{Majumder2015, Baek2017, Wolter2017, Leahy2017, Hentrich2018, Kasahara2018, Banerjee2018}. In addition, recent
high-pressure experiments showed that a tiny hydrostatic pressure of $p\backsimeq 0.2$~GPa induces
a structural transition toward a valence-bond crystal~\cite{Cui2017, Bastien2018}. Chemical substitution of Cl by Br leads to an expansion of the honeycomb lattice and a concomitant increase of the N\'eel temperature~\cite{Huening2001}.

A viable route to tune the magnetic properties of Kitaev magnets and in particular of
$\alpha$-RuCl$_3$ is chemical substitution on the magnetic-ion site.
This has been discussed in several theoretical studies for the pure Kitaev
limit~\cite{Willans2010, Dhochak2010, Willans2011, Vojta2016, Das2016}, where it was shown that
the spin-liquid state is relatively robust under doping and that the Majorana fermions can screen magnetic impurities resulting in a Kondo effect.
Experimental studies on honeycomb iridates showed that the substitution of Ir$^{4+}$ by nonmagnetic ions, such
as Ti$^{4+}$ in Li$_2$Ir$_{1-x}$Ti$_x$O$_3$ and Na$_2$Ir$_{1-x}$Ti$_x$O$_3$, and also by strongly
spin-orbit-coupled magnetic ions, e.g. Ru$^{4+}$ in Na$_2$Ir$_{1-x}$Ru$_x$O$_3$, changes the
antiferromagnetic zigzag state into a spin-glass state~\cite{Manni2014, Mehlawat2015}. Via
Monte Carlo calculations on the Kitaev-Heisenberg model this formation of a spin-glass
state under the substitution of the magnetic ion by a nonmagnetic ion~\cite{Andrade2014} or by
magnetic ions~\cite{Cai2017} has successfully been rationalized.
On the other hand, in $\alpha$-RuCl$_3$ the substitution by nonmagnetic Ir$^{3+}$ did not lead to  a spin-glass state in Ru$_{1-x}$Ir$_x$Cl$_3$. Instead, the N\'eel temperatures corresponding to the regions of ABC and of AB stacking domains in Ru$_{1-x}$Ir$_x$Cl$_3$ both decrease continuously with the Ir content $x$ and vanish around $x\backsimeq 0.11$ and $x\backsimeq 0.24$,
respectively~\cite{Lampen-Kelley2017, Do2018}. The nature of the quantum disordered state beyond
$x\backsimeq 0.24$ is under debate, with a dilute spin liquid being proposed from inelastic
neutron scattering experiments in Ref.~\onlinecite{Lampen-Kelley2017}.



\begin{figure}[t]
\begin{center}
\includegraphics[width=0.9\linewidth]{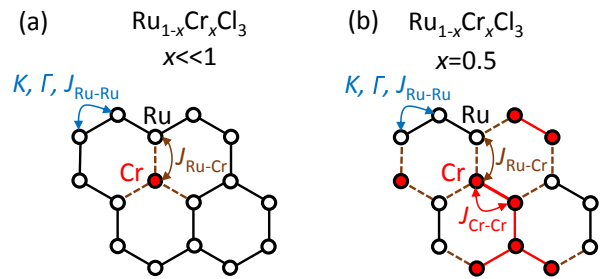}
\caption{(color online)(a) Schematic view of the honeycomb layer built by the Ru and Cr ions in
Ru$_{1-x}$Cr$_x$Cl$_3$ in the $x\ll 1$ limit. The red and white balls represent the Cr and Ru sites,
respectively. The black, brown and red nearest neighbor links correspond to Ru-Ru, Ru-Cr and Cr-Cr
nearest-neighbor links, respectively, together with the arrows indicating the dominant magnetic nearest-neighbor
interactions. (b) Same for $x=0.5$.} \label{honeycomb}
\end{center}
\end{figure}

In this paper we focus on the substitution of Ru$^{3+}$ in $\alpha$-RuCl$_3$ by magnetic
Cr$^{3+}$ ions.The possibility of crystal growth of Ru$_{1-x}$Cr$_x$Cl$_3$ crystals with a homogeneous distribution of Ru and Cr on the same crystallographic site was previously reported~\cite{Hillebrecht1997, Roslova2018}. The magnetic moment of the Cr site is not subject to  strong spin-orbit coupling
and can thus be considered as a spin-only moment $S = 3/2$. The honeycomb layer of
Ru$_{1-x}$Cr$_x$Cl$_3$ is represented in Fig.~\ref{honeycomb} together with the presumed main
magnetic interactions. While the magnetic interactions on Ru-Ru nearest-neighbor links can be
assumed to be the sum of Kitaev, Heisenberg and off-diagonal interactions, the magnetic
interactions on Cr-Cr nearest-neighbor links are expected to be of Heisenberg-type. The nature of
the magnetic interactions on the nearest-neighbor links Ru-Cr, however, remains to be determined.
Thus, for low doping levels $x\ll 1$, the Ru$_{1-x}$Cr$_x$Cl$_3$ series allows to study the role of
magnetic impurities in a Kitaev magnet, while for higher doping levels the evolution
from a frustrated Kitaev magnet to a non-frustrated Heisenberg magnet on a honeycomb lattice can be investigated.



The final member of the series, CrCl$_3$, crystallizes in the same monoclinic crystal structure as
$\alpha$-RuCl$_3$~\cite{Morosin1964}.
It orders antiferromagnetically at $T_N=14~$K with a ferromagnetic in-plane alignment of the spins
within the honeycomb layers and an antiparallel alignment of the spins for adjacent
layers~\cite{Cable1961, McGuire2017}. Further, a two-dimensional ferromagnetic ordering at $T_{2D}
= 17$~K prior to the long-range antiferromagnetic ordering at 14~K was proposed from Faraday
rotation and specific-heat measurements~\cite{Kuhlow1982, McGuire2017}.

It is the purpose of this paper to discuss the magnetic properties of the complete Ru$_{1-x}$Cr$_x$Cl$_3$ series based on magnetization, specific-heat, and ac susceptibility measurements. The combination of these thermodynamic probes allows us to construct the $x$--$T$ phase diagram of Ru$_{1-x}$Cr$_x$Cl$_3$, where the antiferromagnetic zigzag order is replaced by a spin-glass state for $0.1\lesssim x \lesssim 0.7$. Interestingly, the evolution of the magnetic anisotropy in the Ru$_{1-x}$Cr$_x$Cl$_3$ series reveals a competition between anisotropic interactions on the Ru-Ru links and more isotropic interactions on Ru-Cr and Cr-Cr links. Notably, the substitution series Ru$_{1-x}$Cr$_x$Cl$_3$ constitutes one of the rare cases with simultaneous tuning of frustration and disorder.

\section{Experimental techniques}

The Ru$_{1-x}$Cr$_x$Cl$_3$ single crystals were grown by a chemical vapor transport reaction in a
two-zone furnace with a temperature gradient between 750~$^\circ$C and 650~$^\circ$C for 5 days.
X-ray diffraction, transmission electron microscopy and Raman spectroscopy showed, that the
monoclinic crystal structure of $\alpha$-RuCl$_3$ is preserved under an homogeneous substitution of
Ru by Cr atoms in the whole series, despite an enhancement of stacking disorder under
doping~\cite{Roslova2018}. The Ru:Cr ratio of the crystals used for our physical properties
measurements was additionally confirmed by energy-dispersive X-ray spectroscopy (EDX). The oxidation state of Ru and Cr can reasonably be assumed to be 3+ in the whole series like in $\alpha$-RuCl$_3$ and CrCl$_3$. The  formation of Cr$^{2+}$ and Ru$^{4+}$ for particular Cr content values cannot be fully excluded, however, it would lead to a Jahn-Teller distortion, which was not observed in previous X-ray diffraction and Raman spectroscopy experiments~\cite{Roslova2018}.
RhCl$_3$ single crystals were used as a nonmagnetic
structural analog compound for the analysis of the specific heat. They were also grown by chemical
vapor transport starting from a Rh metal powder and Cl$_2$ gas in the ratio $n($Cl$_2)/n($Rh) = 1.6
under a temperature gradient from $900~^\circ$C to $800~^\circ$C during six days. The absence of
(magnetic) impurity phases was confirmed.

Dc and ac magnetic susceptibility and specific-heat measurements were conducted with a commercial
superconducting quantum interference device (SQUID) magnetometer MPMS-XL and a physical property
measurement system (PPMS) by Quantum Design, respectively. For the specific-heat studies a
heat-pulse relaxation technique was used. For each measurement of the dc magnetization and the
specific heat, the background signal of the sample holder was measured separately and subtracted
from the total raw signal.

\section{DC Magnetization measurements}

\subsection{Temperature dependence of the magnetization}

\begin{figure}[t]
\begin{center}
\includegraphics[width=0.9\linewidth]{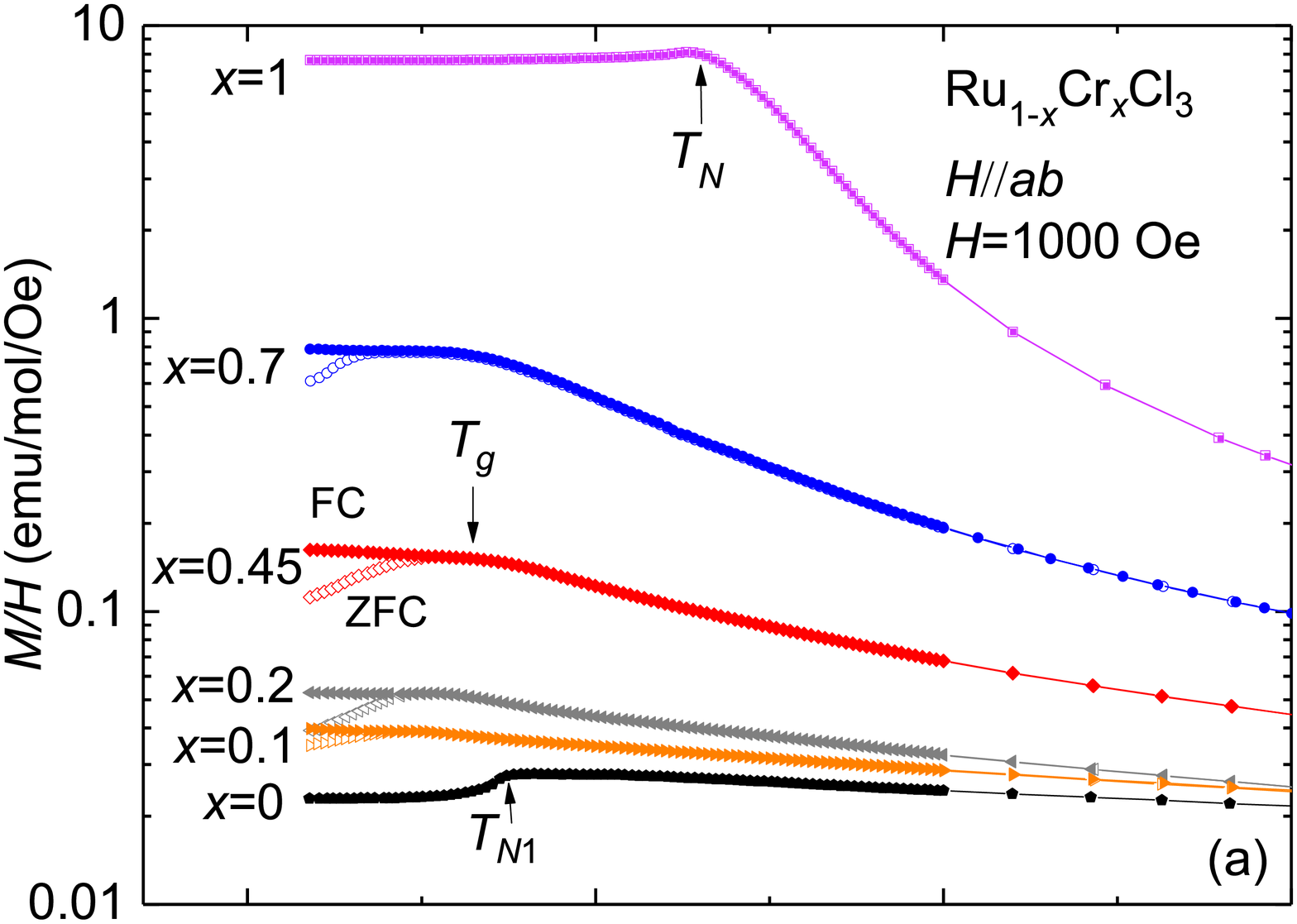}
\includegraphics[width=0.9\linewidth]{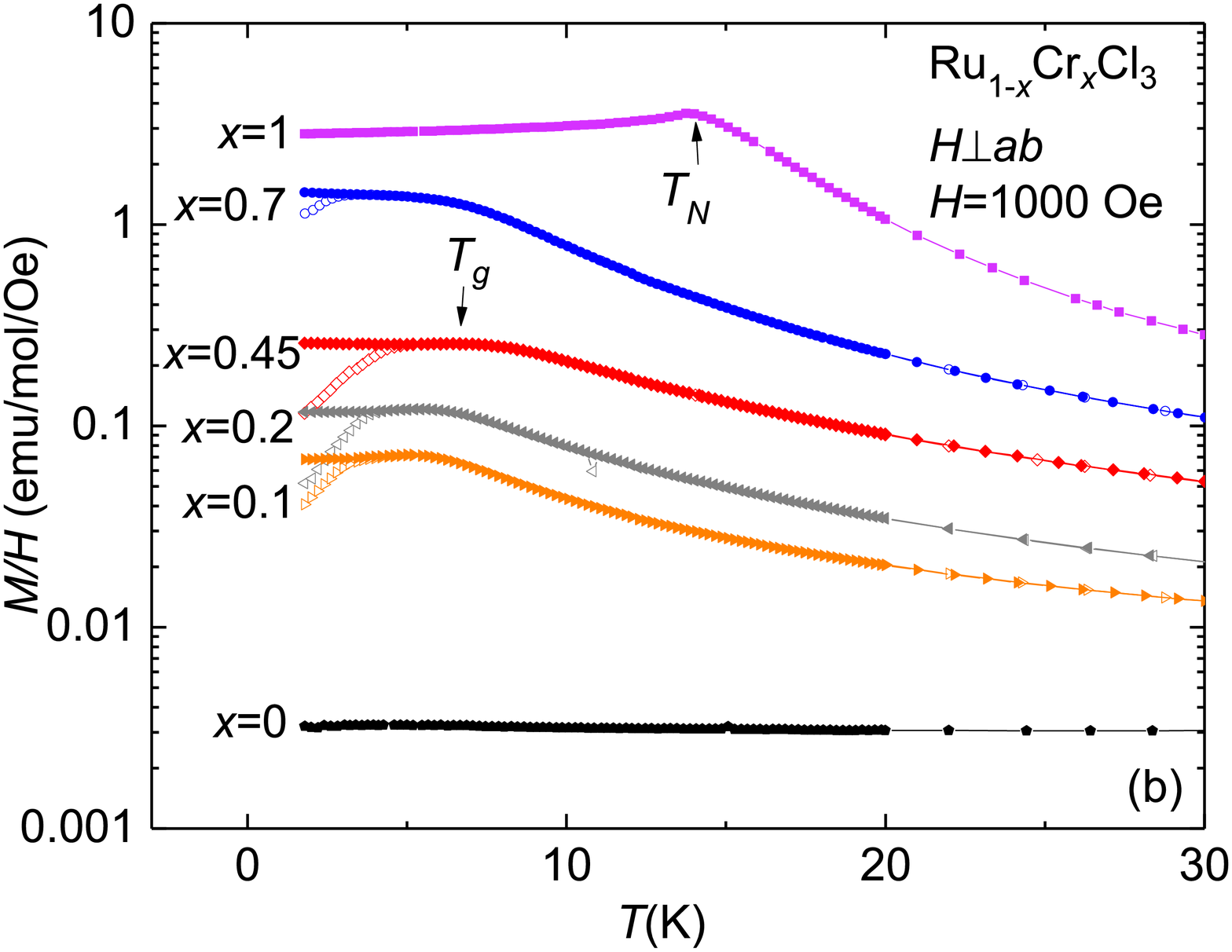}
\caption{(color online)(a) Magnetization divided by magnetic field $M/H$  of
Ru$_{1-x}$Cr$_x$Cl$_3$ in the $ab$ plane on a logarithmic scale as a function of temperature. No demagnetization correction has been applied. Open and full symbols
stand respectively for zero-field-cooled and field-cooled magnetization measurements. The magnetic transition temperatures $T_N$ and
$T_g$ are indicated by a change of slope of the
magnetization as function of temperature. (b) Same figure but for a transverse field: $H\bot ab$.}
\label{MvsT}
\end{center}
\end{figure}

The temperature dependence of the magnetization up to 30~K of Ru$_{1-x}$Cr$_x$Cl$_3$ is represented
on a log scale in Fig.~\ref{MvsT} for a magnetic field $\mu_0H$ = 0.1~T applied in an arbitrary direction within the $ab$ plane and
transverse to the $ab$ plane \footnote{The possible occurrence of an in-plane magnetic anisotropy in Ru$_{1-x}$Cr$_x$Cl$_3$ was not investigated in the present study, but is considered to get smaller upon Cr doping anyhow.}.
For both field directions the progressive increase on two orders of
magnitude of the low-temperature magnetization with doping suggests the appearance of ferromagnetic
interactions between Cr atoms in the basal plane: $J_{Cr-Cr} < 0$. For doping levels in the range
$0.1 \leq x \leq 0.7$ and for both field directions a phase transition can be observed via a change
of slope at $T_g\approx$ 5-7~K followed by a splitting between the zero-field-cooled (ZFC) and
field-cooled (FC) magnetization curves at low temperature. Finally, the magnetization of CrCl$_3$ ($x = 1$) shows an antiferromagnetic transition for both field directions in good agreement with
previous magnetization studies on this material~\cite{Bizette1961, McGuire2017}. This transition is
strongly field dependent~\cite{McGuire2017} and its extrapolation to zero field gives $T_N=13.9~$K
for our crystals in agreement with Ref.~\onlinecite{McGuire2017}.

\begin{figure*}[t]
\begin{minipage}{0.45\linewidth}
\begin{center}
\includegraphics[width=0.98\linewidth]{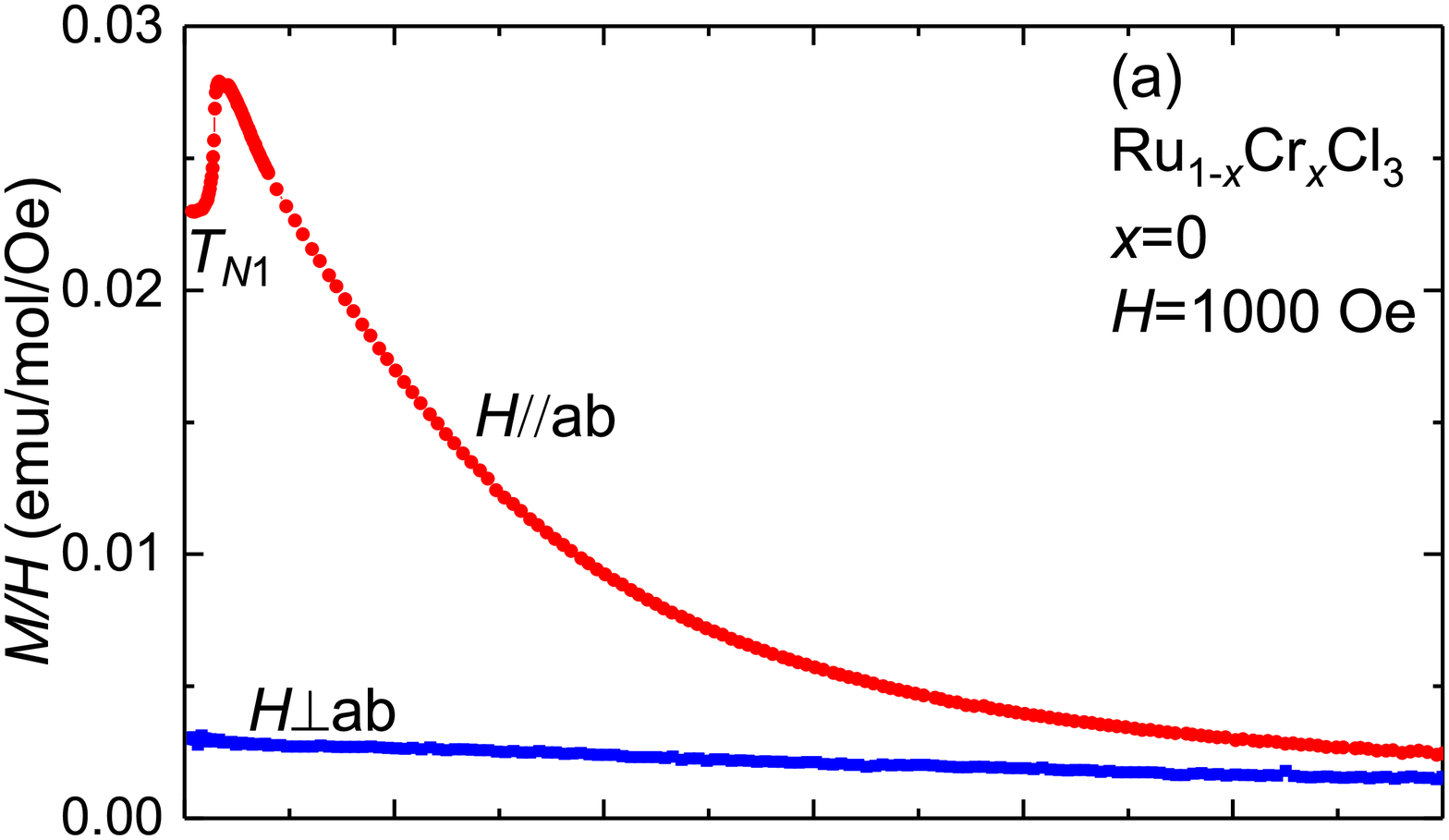}
\includegraphics[width=0.98\linewidth]{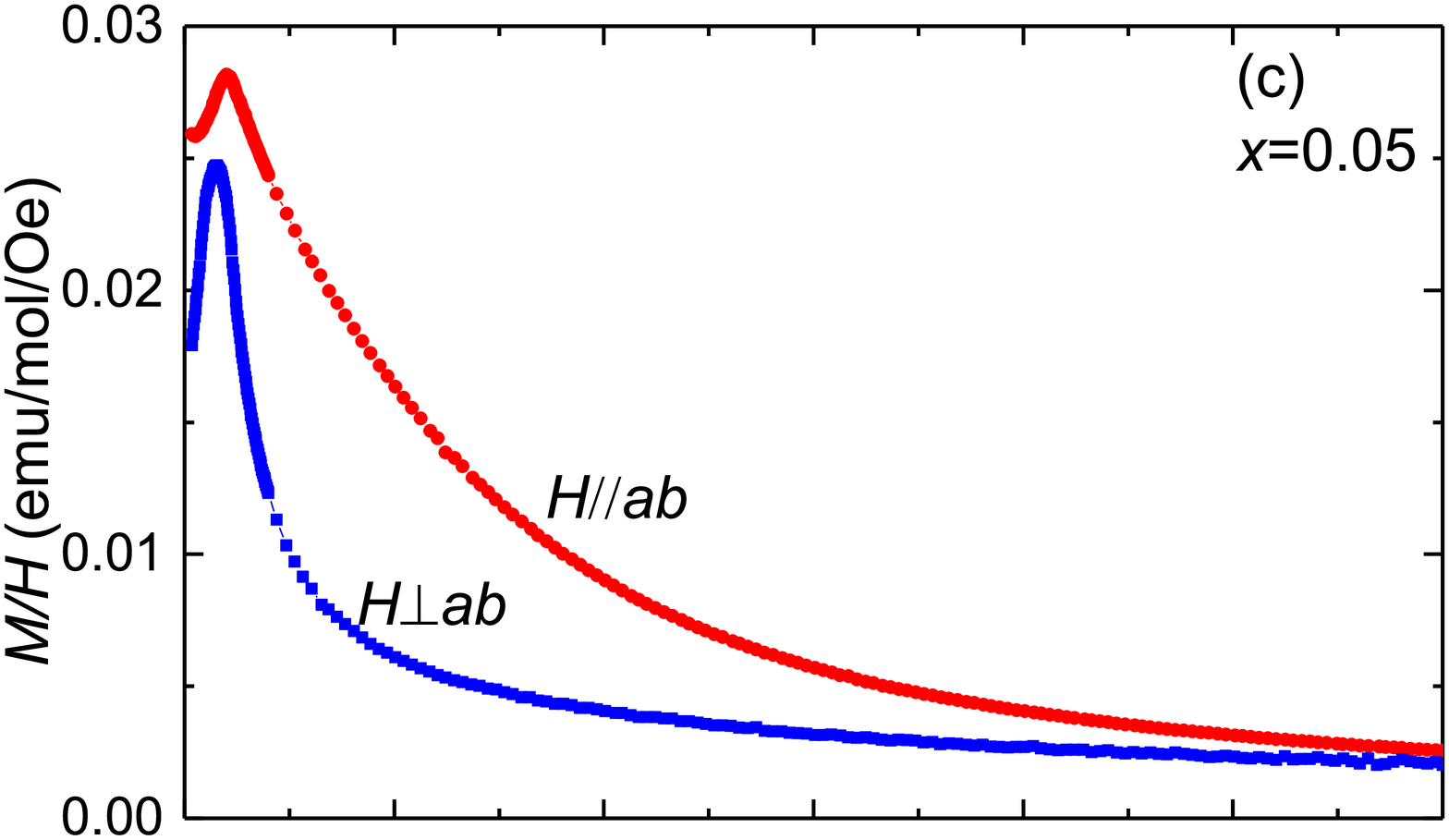}
\includegraphics[width=0.98\linewidth]{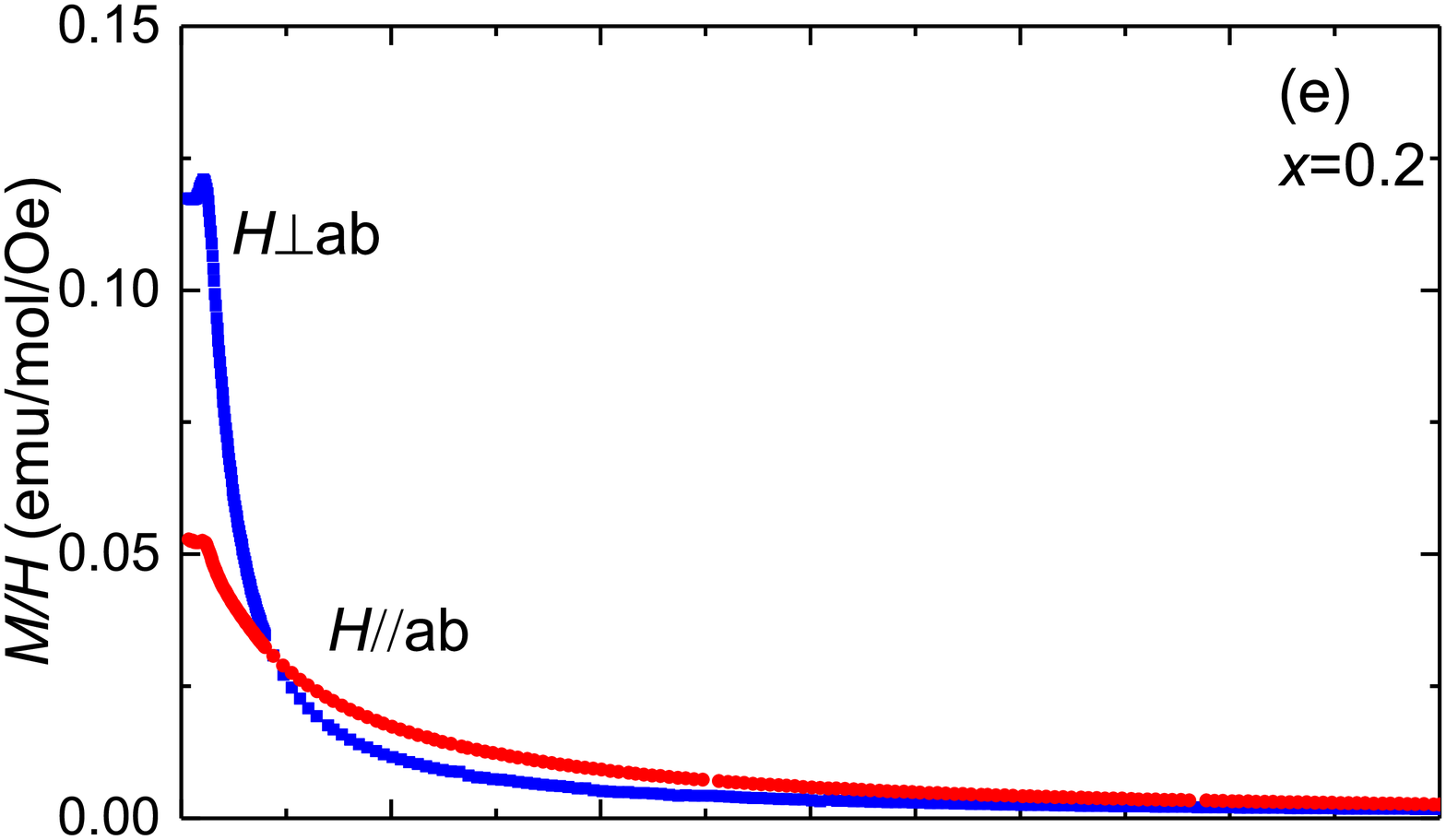}
\includegraphics[width=0.98\linewidth]{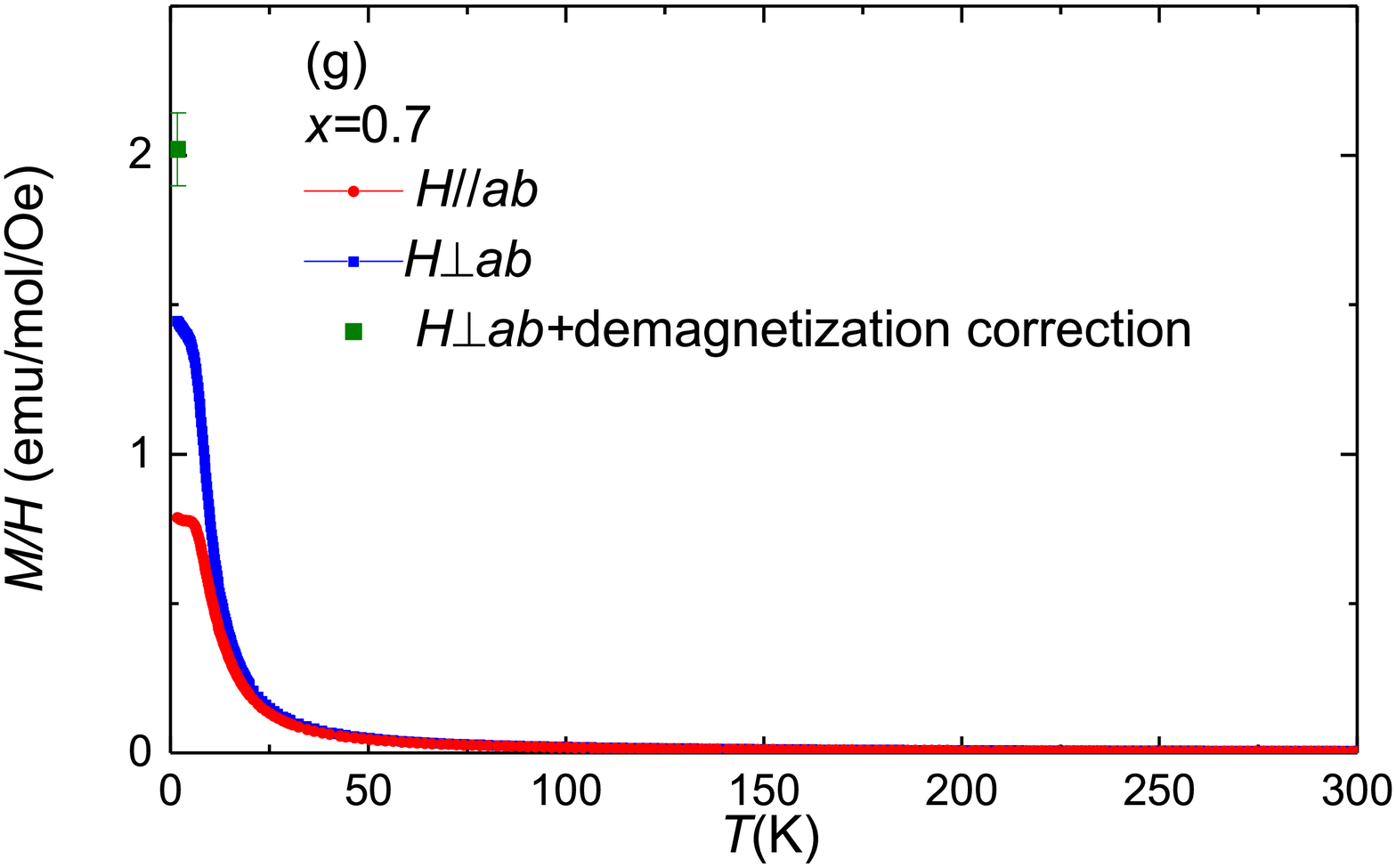}
\end{center}
\end{minipage}
\hfill
\begin{minipage}{0.45\linewidth}
\begin{center}
\includegraphics[width=0.98\linewidth]{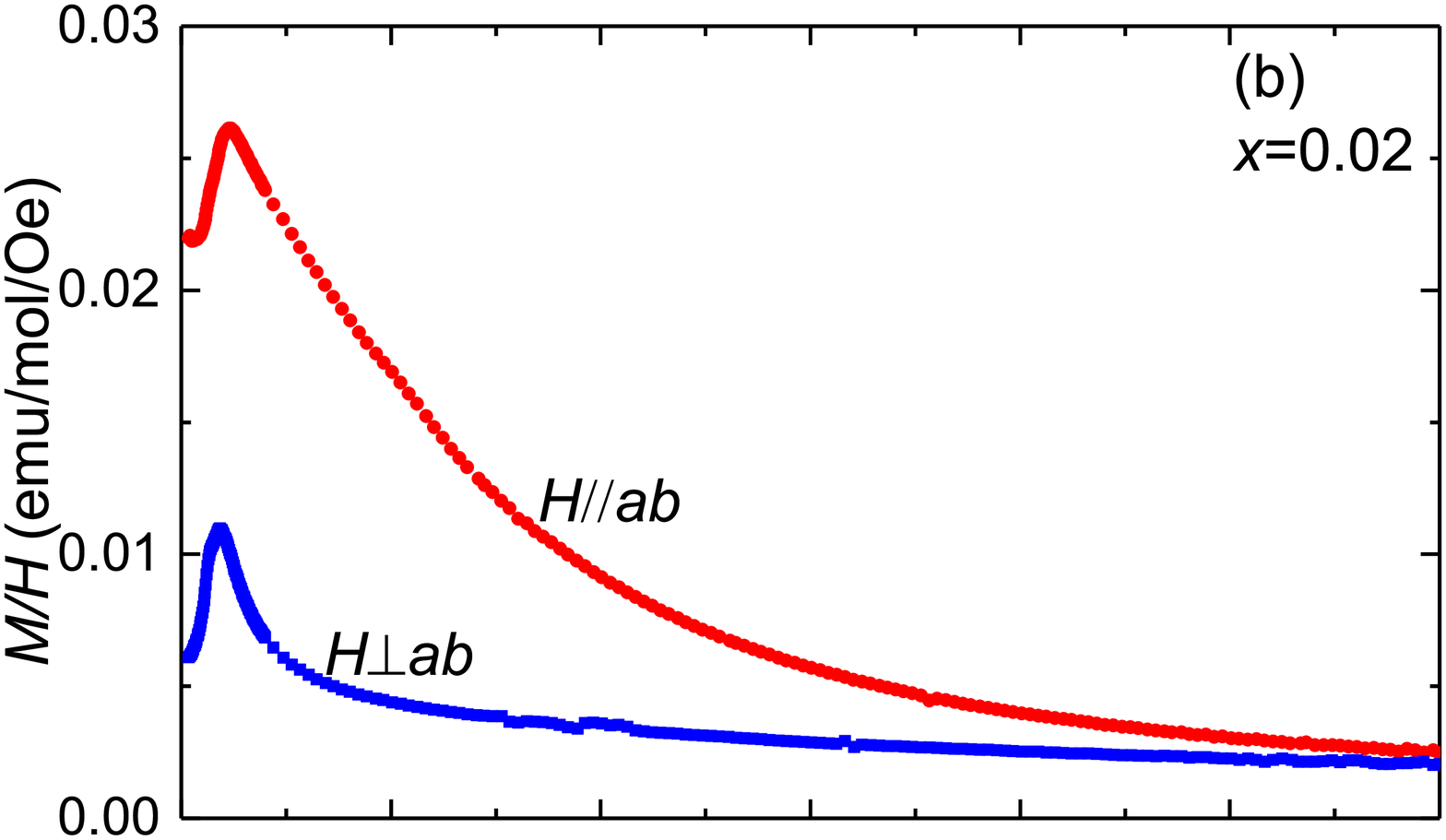}
\includegraphics[width=0.98\linewidth]{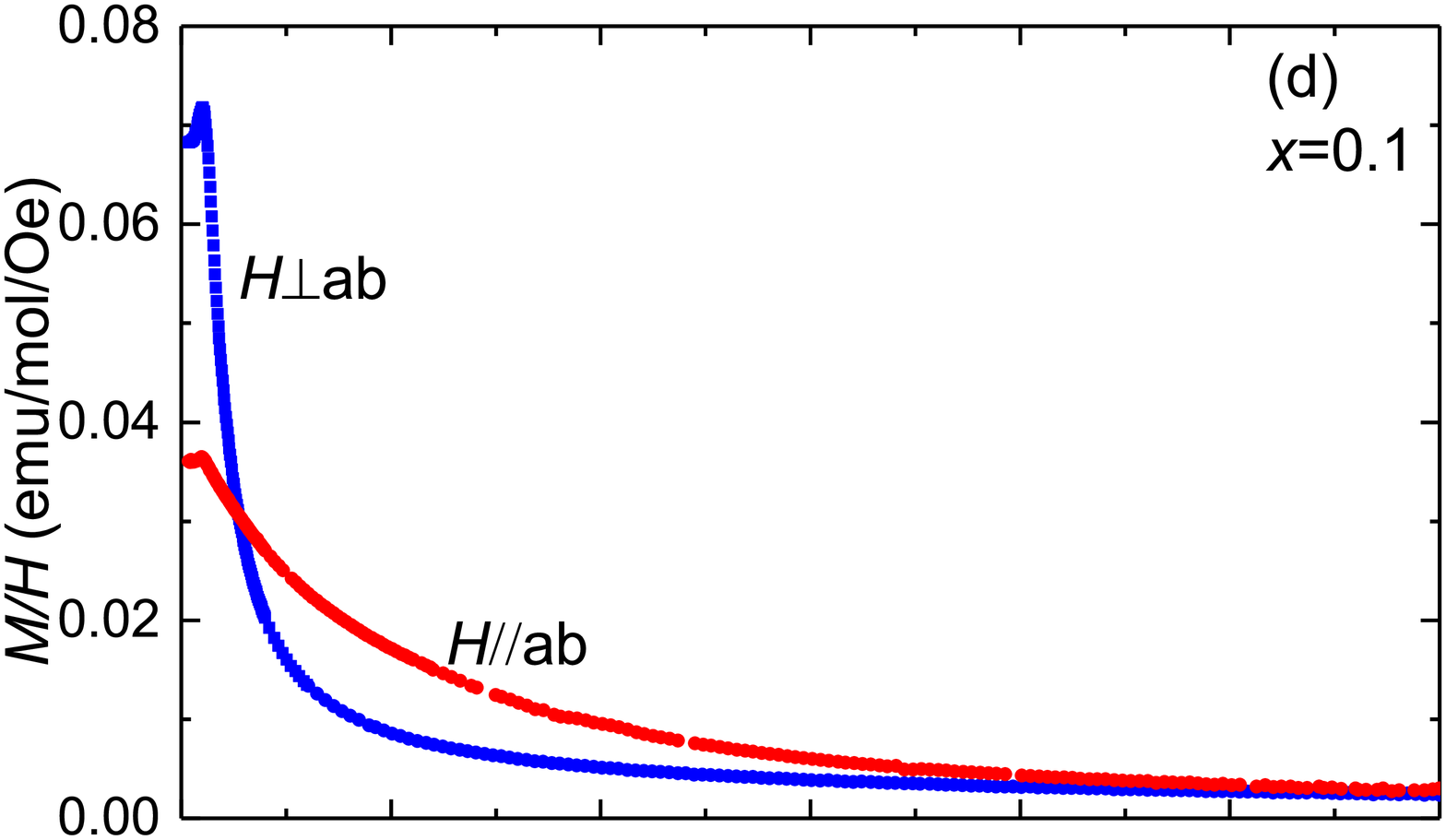}
\includegraphics[width=0.98\linewidth]{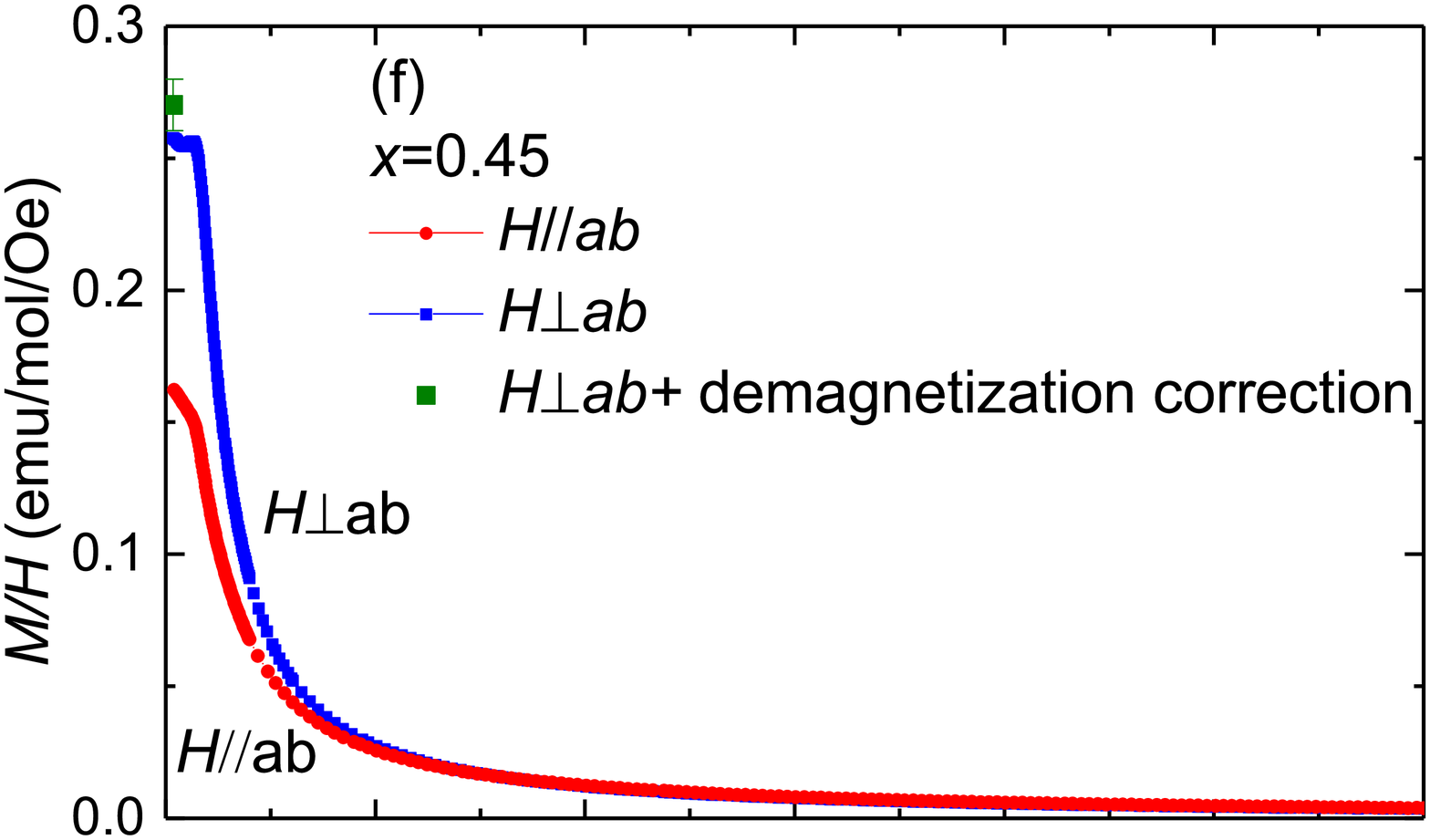}
\includegraphics[width=0.98\linewidth]{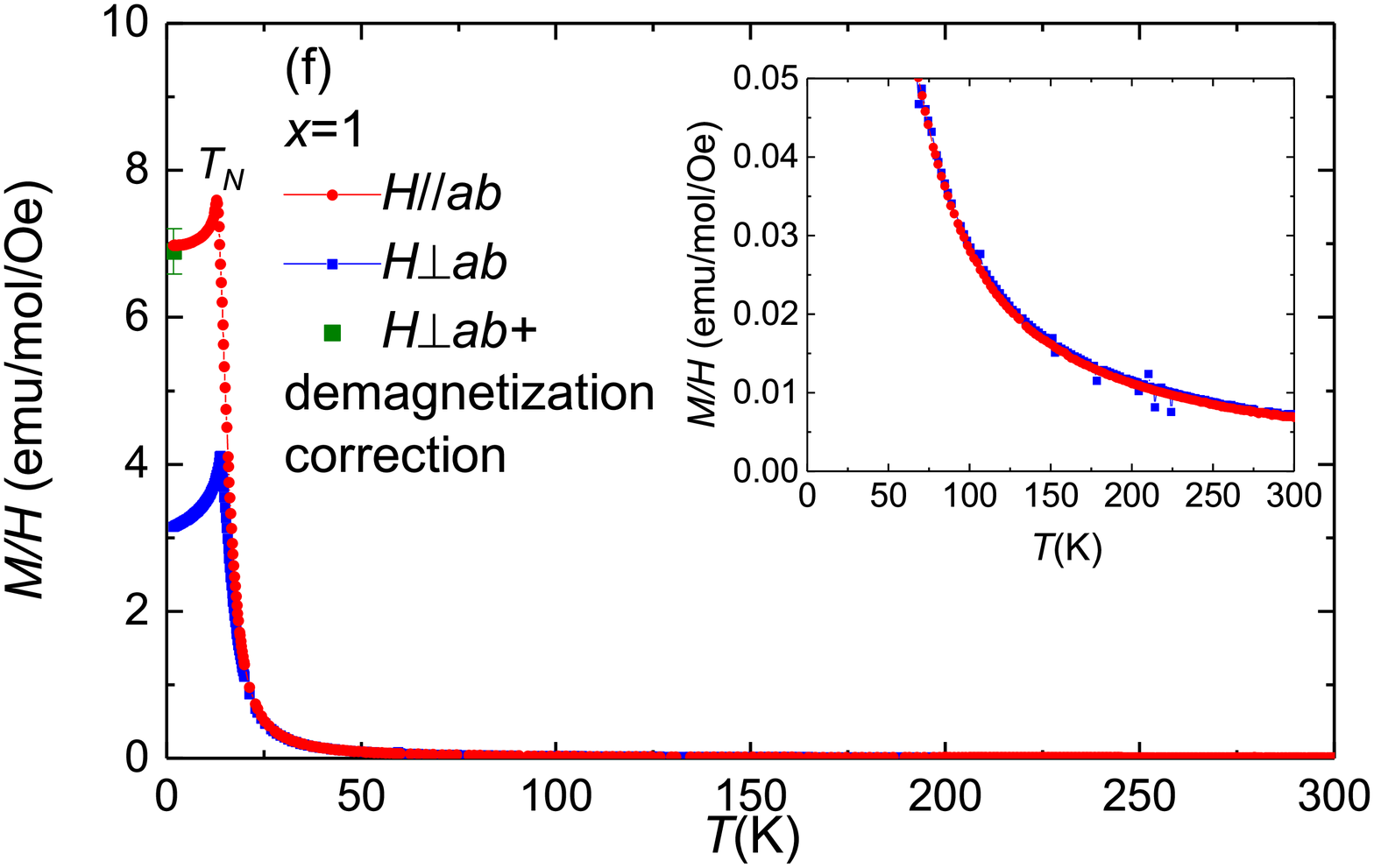}
\end{center}
\end{minipage}
\caption{
 (color online) Magnetization divided by the magnetic field $M/H$ of Ru$_{1-x}$Cr$_x$Cl$_3$ for magnetic field applied in the $ab$
plane and transverse to the $ab$ plane for (a) $x=0$, (b) $x=0.02$,
(c) $x=0.05$, (d) $x=0.1$, (e) $x=0.2$, (f) $x=0.45$, (g) $x=0.7$
(h) $x=1$. The green points at $T$=1.8~K in (f), (g) and (h) represent the magnetization transverse
to the $ab$ plane after correction of the demagnetization effect. This correction was computed from
the field dependence of the magnetization represented in Fig.~\ref{MvsH} as detailed below. 
For
CrCl$_3$ ($x=1$) the magnetization at $T$=1.8~K after correction of the demagnetization effect does not
show any anisotropy within the error bars.
}
\label{MvsTanisotropy}
\end{figure*}

The magnetization curves of some Ru$_{1-x}$Cr$_x$Cl$_3$ single crystals as a function of
temperature extending up to 300~K are represented in Fig.~\ref{MvsTanisotropy} for both magnetic
fields applied in the basal plane $ab$ and transverse to it. For $\alpha$-RuCl$_3$ ($x$ = 0) the
strong magnetic anisotropy previously reported is reproduced~\cite{Majumder2015}. This magnetic
anisotropy is a signature of the strong spin-orbit coupling on the Ru site and has mainly been
related to the occurrence of off-diagonal $\Gamma$ magnetic interactions~\cite{Lampen-Kelley2018}. 

Following the substitution series to amounts of $x=0.02$ and $x=0.05$ substitution of Ru$^{3+}$ by Cr$^{3+}$,
a strong increase of the magnetic susceptibility transverse to the basal plane $ab$ is observed
already in the paramagnetic state, without any clear increase of the in-plane magnetic
susceptibility. For higher Cr contents $x \geq 0.1$ the magnetic anisotropy gets even reversed upon cooling, with a higher magnetic susceptibility transverse to the basal plane at low temperatures $T \lesssim$ 15~K. Such changes of the magnetic anisotropy with temperature are not very common.
It was observed in the spin-chain magnet Sr$_3$NiIrO$_6$ and explained from the competition between the single-ion anisotropy at the Ni site and anisotropic interactions between Ni and Ir moments~\cite{Lefrancois2016}. 
In Ru$_{1-x}$Cr$_x$Cl$_3$ the change of the magnetic anisotropy with temperature can probably be
attributed to different anisotropies of the magnetization contributions from Ru and Cr moments, see below for a discussion. For an even higher Cr content $x=0.45$, the magnetic anisotropy is
reversed compared to the one of $\alpha$-RuCl$_3$ for the full temperature range, however,
collapses at high temperature ($T > $ 150~K). Thus the magnetic anisotropy
for $x=0.45$ is opposite to the expected anisotropy of the Ru moments given by
the $\Gamma$ interaction on the Ru-Ru links. Finally, for the binary compound CrCl$_3$($x = 1$), a negligible magnetic anisotropy is observed after the consideration of demagnetization effects (see
Fig.~\ref{MvsTanisotropy} and Ref.~\onlinecite{McGuire2017}), evidencing the small single-ion anisotropy of Cr moments together with the small anisotropy for Cr-Cr magnetic exchange interactions.

\begin{figure}[t]
\begin{center}
\includegraphics[width=0.9\linewidth]{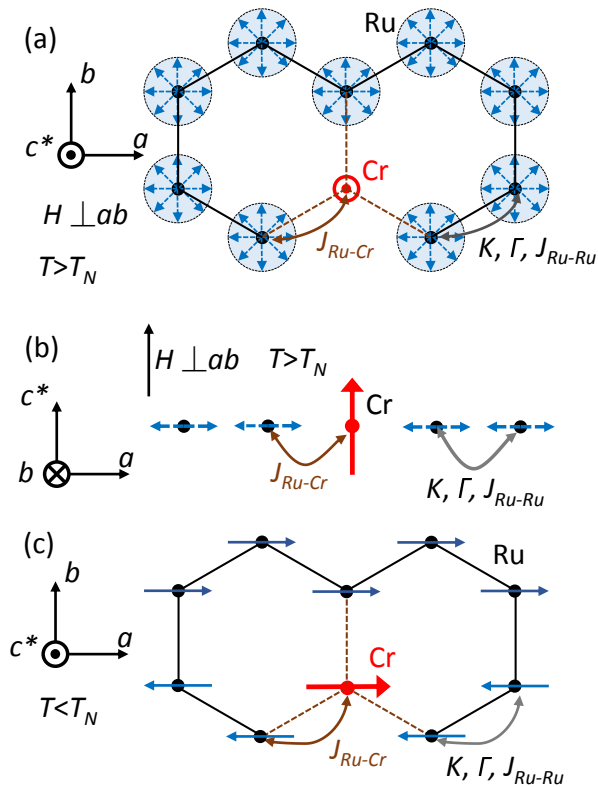}
\caption{(color online)(a) Schematic view of the magnetic properties of Ru$_{1-x}$Cr$_x$Cl$_3$ in the low Cr content limit $x\ll 1$, in the paramagnetic state $T>T_N$ and under a magnetic field applied perpendicular to the $ab$ plane, i.e., along the $c^*$ axis. The collective alignment of the moments on the Ru site toward the field is prevented by the off-diagonal interaction $\Gamma$. As a simplified picture, the moments on the Ru site are represented by blue arrows laying in the basal plane $ab$. The solid red arrow indicates the spin of the Cr ion, which is polarized by the field transverse to the basal plane. (b) Schematic view from the $b$ axis under the same conditions. (c) Schematic view of the magnetic properties of Ru$_{1-x}$Cr$_x$Cl$_3$ in the low Cr content limit $x\ll 1$, but now in the ordered state $T<T_N$. The red and blue arrows represent the magnetic moments on the Cr and Ru sites respectively. They are represented in the figure assuming an isotropic antiferromagnetic Ru-Cr interaction. The tilting of the spins out of the basal plane~\cite{Cao2016} is not represented here for simplicity. The ordering of Ru and Cr magnetic moments explain the reduction of both the magnetic susceptibility in the $ab$ plane and transverse to the $ab$ plane.
}
\label{scheme}
\end{center}
\end{figure}

In order to interpret these findings, we now sketch the most likely scenario for the
underlying magnetic exchange in Ru$_{1-x}$Cr$_x$Cl$_3$, focusing on the response in the
paramagnetic region. 
The different couplings $K$, $\Gamma$, $J_{Ru-Ru}$, $J_{Ru-Cr}$ and $J_{Cr-Cr}$  must evolve with doping, since they are sensitive to the bond angles~\cite{Yadav2016}. In the following, we assume that none of these coupling parameters vanishes or changes the sign upon substitution (0$\leq x \leq$ 1) and the consistency of the data within this assumption will be discussed.
First of all, the absence of a clear increase of the in-plane magnetization for small values $x \leq 0.05$ indicates an antiferromagnetic nature of the Ru-Cr magnetic exchange interactions.
For simplicity, this interaction will be assumed to be of Heisenberg-type. Then, the strong
increase of the out-of-plane magnetization under doping must stem from the magnetic moment on the
Cr site ($S$ = 3/2), for which an isotropic moment without sizable single-ion anisotropy can
be expected. Indeed, while a uniform canting of the Ru moments by an out-of-plane field is suppressed by off-diagonal $\Gamma$ interactions on Ru-Ru bonds, the Zeeman effect on the Cr sites
would be able to overcome the Ru-Cr magnetic interactions thanks to the strong magnetic frustration in Ru$_{1-x}$Cr$_x$Cl$_3$. This leads to an out-of-plane alignment of the Cr spins for fields
transverse to $ab$ in the paramagnetic regime as schematically depicted in Fig.~\ref{scheme}.

\begin{figure}[t]
\begin{center}
\includegraphics[width=0.9\linewidth]{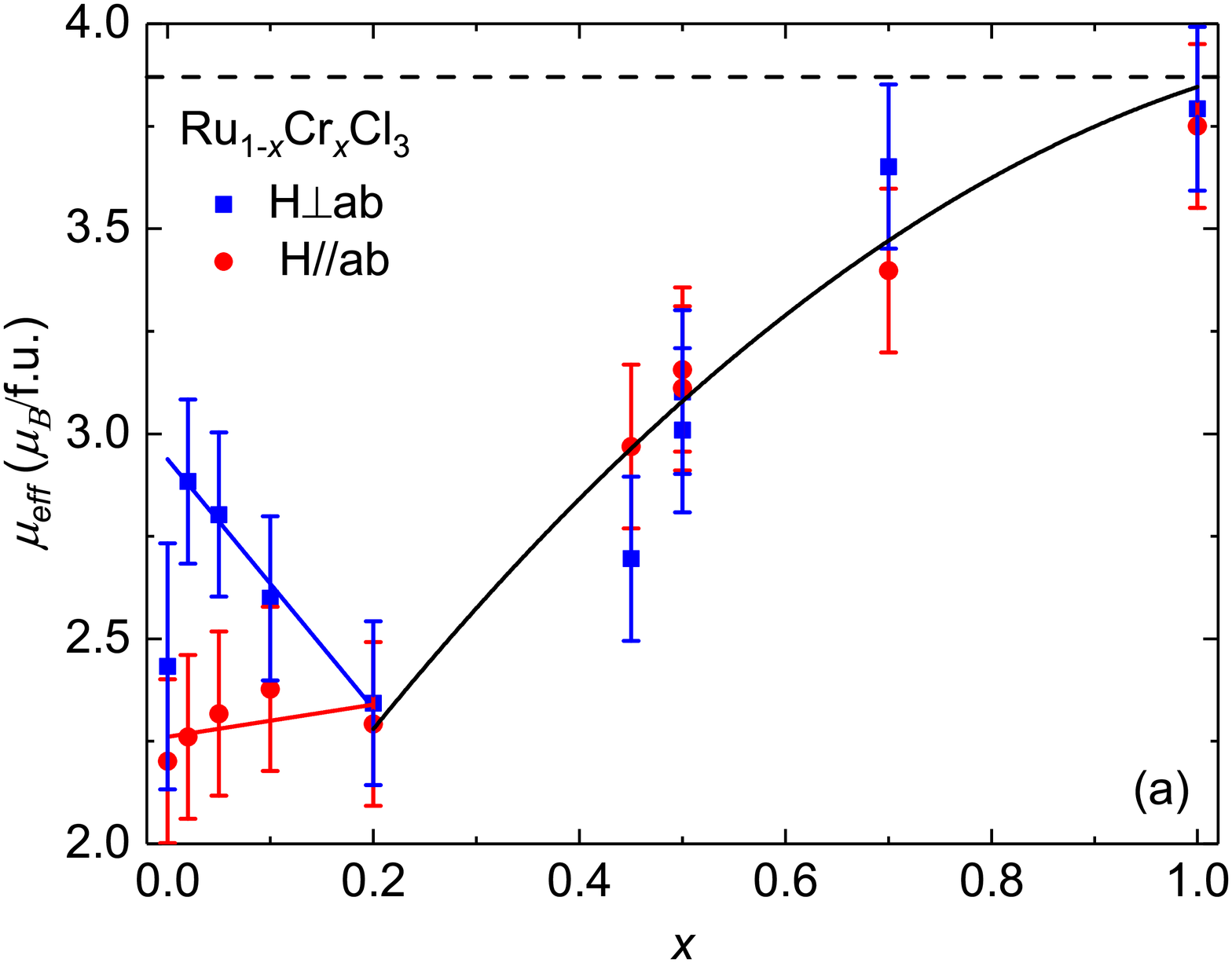}
\includegraphics[width=0.9\linewidth]{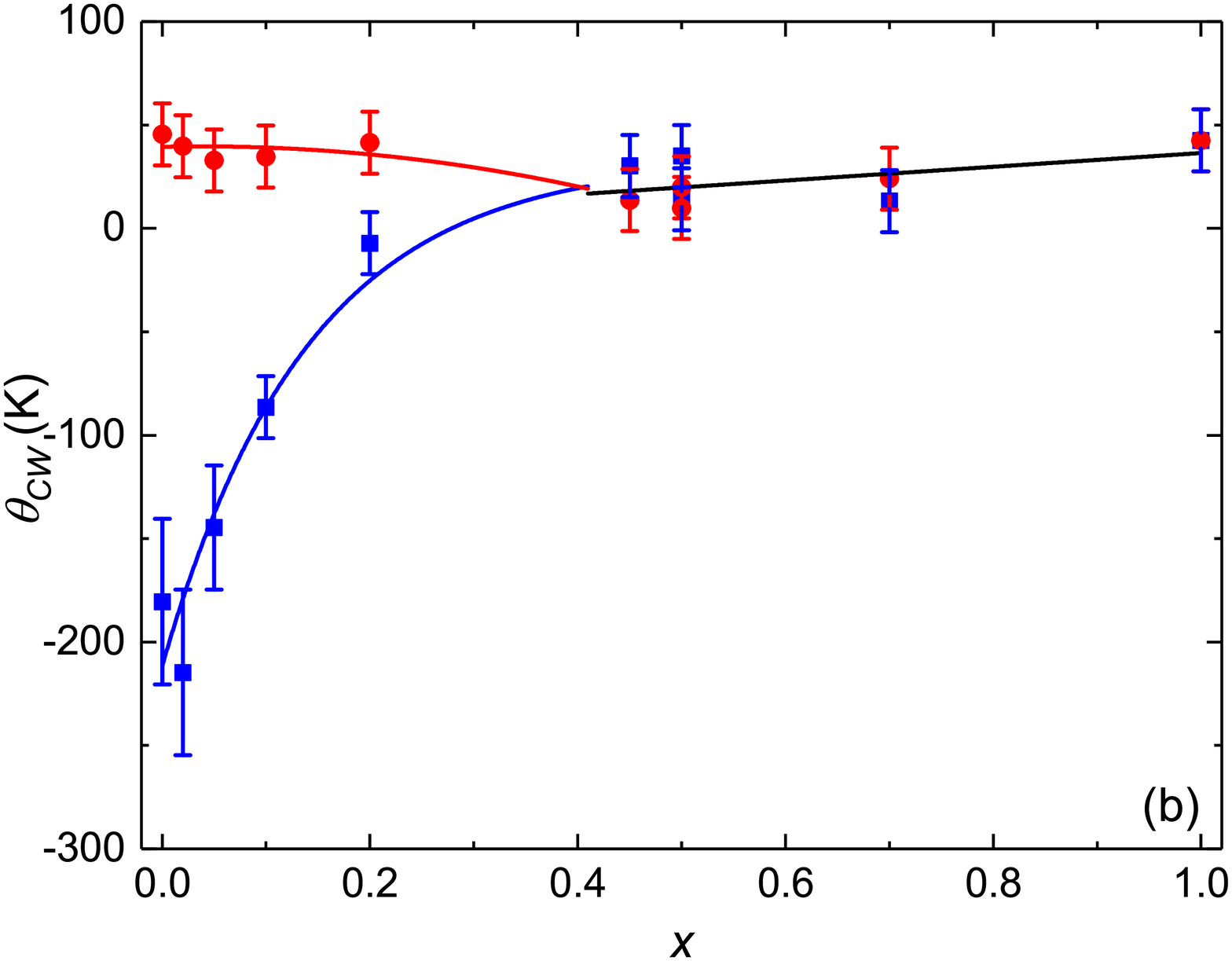}
\caption{(color online)(a) Averaged effective moment of the magnetic ion Ru/Cr for Ru$_{1-x}$Cr$_x$Cl$_3$ as a
function of the Cr content $x$ for magnetic fields applied in the $ab$ plane and transverse to the
$ab$ plane. Curie-Weiss fits were performed in the temperature interval 200~K$ < T < $300~K at
$\mu_0H = 1$~T. The solid lines are guides to the eye, a single line was drawn above $x = $0.2,
where the anisotropy of the effective moment is smaller than the resolution of the measurement. The
dashed line indicates the free-ion effective moment of Cr$^{3+}$. (b) Fitted Curie-Weiss
temperatures for Ru$_{1-x}$Cr$_x$Cl$_3$ as a function of the Cr content $x$ for magnetic fields
applied in the $ab$ plane and transverse to the $ab$ plane.} \label{CW}
\end{center}
\end{figure}

In this scenario, the reversed magnetic anisotropy for higher Cr dopings, such as e.g. $x$ = 0.45,
can then be understood by the competition between the different magnetic exchange interactions,
which need to be considered in this doping region, i.e, $K$, $\Gamma$, $J_{Ru-Ru}$ on Ru-Ru links,
together with an antiferromagnetic $J_{Ru-Cr}$ and ferromagnetic $J_{Cr-Cr}$ on Ru-Cr and Cr-Cr
links, respectively (see above). Under magnetic fields applied in the $ab$ plane the ferromagnetic Cr-Cr interactions compete with the antiferromagnetic Cr-Ru and the anisotropic Ru-Ru interactions leading to a relatively small increase of the
magnetization despite the presence of ferromagnetic interactions. However, under a magnetic field applied transverse to the $ab$ plane, the alignment of the Ru moments along the field direction is prevented by the anisotropic $\Gamma$ interaction. Thus the Cr spins aligned with the field are less "prone" to existing antiferromagnetic Cr-Ru interactions of Heisenberg-type. This effect leads to an opposite magnetic anisotropy for the contribution of the Ru and of the Cr moments to the magnetization. Then, the resulting measured (total) anisotropy
 is given by the superposition of the two contributions, where the
strongest contribution from Cr moments for $x=0.45$ determines the overall magnetic anisotropy of the system.



The evolution of the magnetic anisotropy in the high temperature limit was mapped out via
Curie-Weiss fittings of the magnetization curves in the temperature interval 200~K
$< T < $300~K, using the equation: $1/\chi=(T-\theta_{CW})/C$ with $C={N\mu_0\mu_{\rm eff}^2}/{3k_B}$.
$N$, $\mu_0$ and $k_B$ stand for the total number of magnetic ions of Ru and Cr, the vacuum
permeability and the Boltzmann constant, respectively. The effective magnetic moments $\mu_{\rm eff}$
and the Curie-Weiss temperatures $\theta_{CW}$ obtained from the fits are represented in
Fig.~\ref{CW} as a function of the Cr content $x$ for both magnetic fields applied parallel and
transverse to the $ab$ plane. The average effective magnetic moment is anisotropic below $x \backsimeq 0.2$: While the effective moment for $H//ab$ is constant within the error bars, the magnetic moment for $H\bot ab$ decreases upon the substitution. Above $x \backsimeq 0.2$, the effective moment is isotropic within the experimental resolution and increases with the Cr content toward $\mu_{\rm eff}=3.8(1)\mu_{B}$/f.u. for CrCl$_3$,
 which is in agreement with the free-ion effective moment of Cr$^{3+}$ of $3.87 \mu_{B}$. The average effective moments of these substitution series are consistent with the assumption of a progressive substitution of Ru$^{3+}$ ions by Cr$^{3+}$ ions without a formation of higher spin ions, i.e., Ru$^{4+}$ with $S=1$ and Cr$^{2+}$ with $S=2$.

The strong anisotropy of the Curie-Weiss temperature of $\alpha$-RuCl$_3$
is a signature of the strong spin-orbit coupling and is related to the occurrence of the
off-diagonal magnetic interactions~\cite{Majumder2015, Lampen-Kelley2018}. The Curie-Weiss
temperature for magnetic fields applied perpendicular to the basal plane decreases in absolute values with the
Cr content from $\theta_{CW} \backsimeq -200 K$ for $x = 0$ and changes its sign to reach $\theta_{CW} \backsimeq
20$~K for $x = 0.45$. The change of behavior of both the effective moment and the Curie Weiss temperature around $x \backsimeq 0.2$ suggests, that the contribution from the moment on the Cr sites dominates the magnetization in the high-temperature regime above $x \backsimeq 0.2$.
Thus, the isotropy of $\mu_{\rm eff}$ and $\theta_{CW}$ for 
$x \geq 0.45$ 
indicates the
absence of strong single-ion anisotropy on the Cr site as well as the existence of dominant isotropic
(Heisenberg-type) Cr-Cr magnetic interactions. The magnetic moment on the Ru site has certainly a
remaining anisotropy due to anisotropic Ru-Ru interactions for $x < 1$, however, its contribution
is too small to give a sizable anisotropy of $\mu_{\rm eff}$ and $\theta_{CW}$ within our
experimental resolution.




\begin{figure}[t]
\begin{center}
\includegraphics[width=0.9\linewidth]{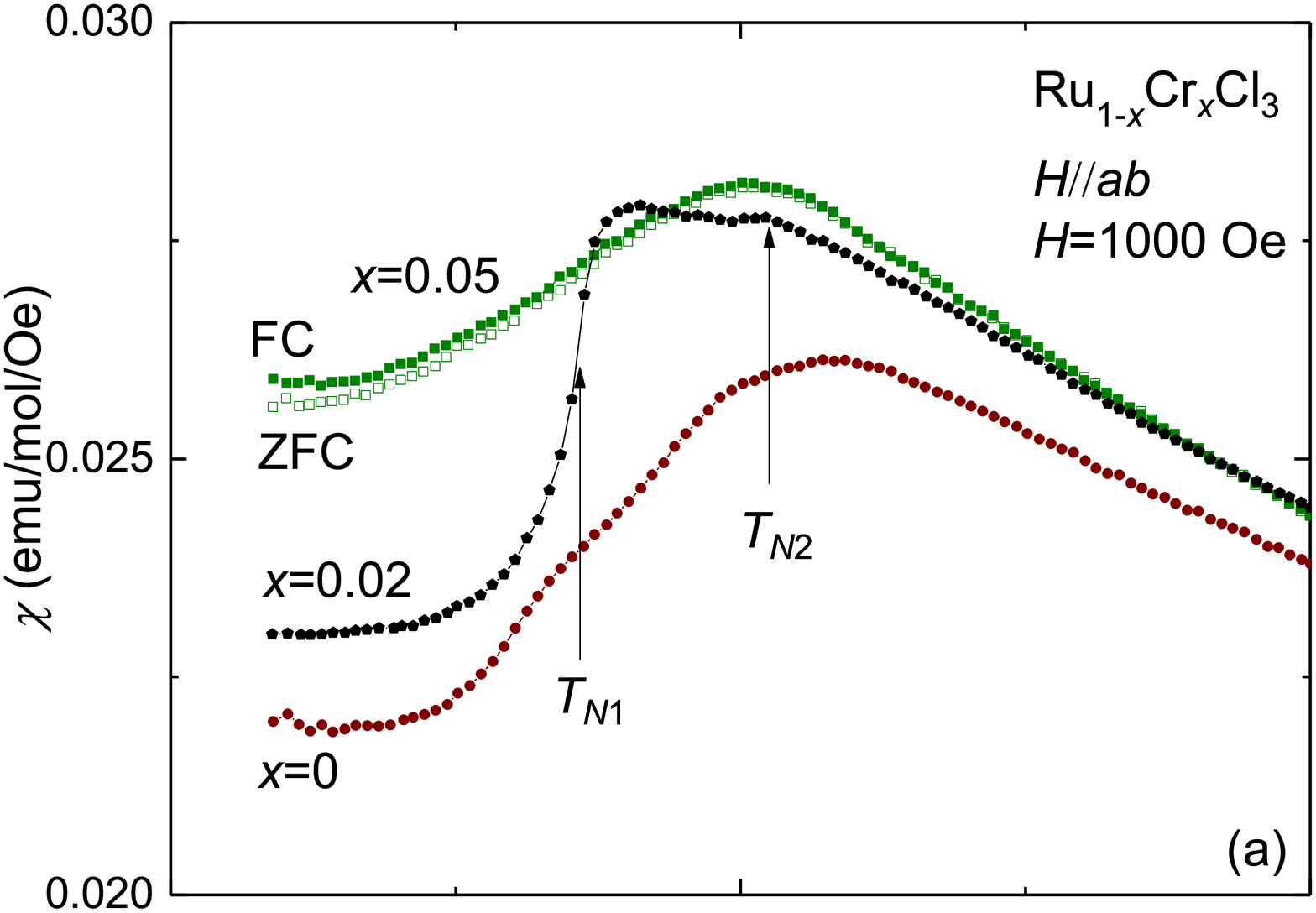}
\includegraphics[width=0.9\linewidth]{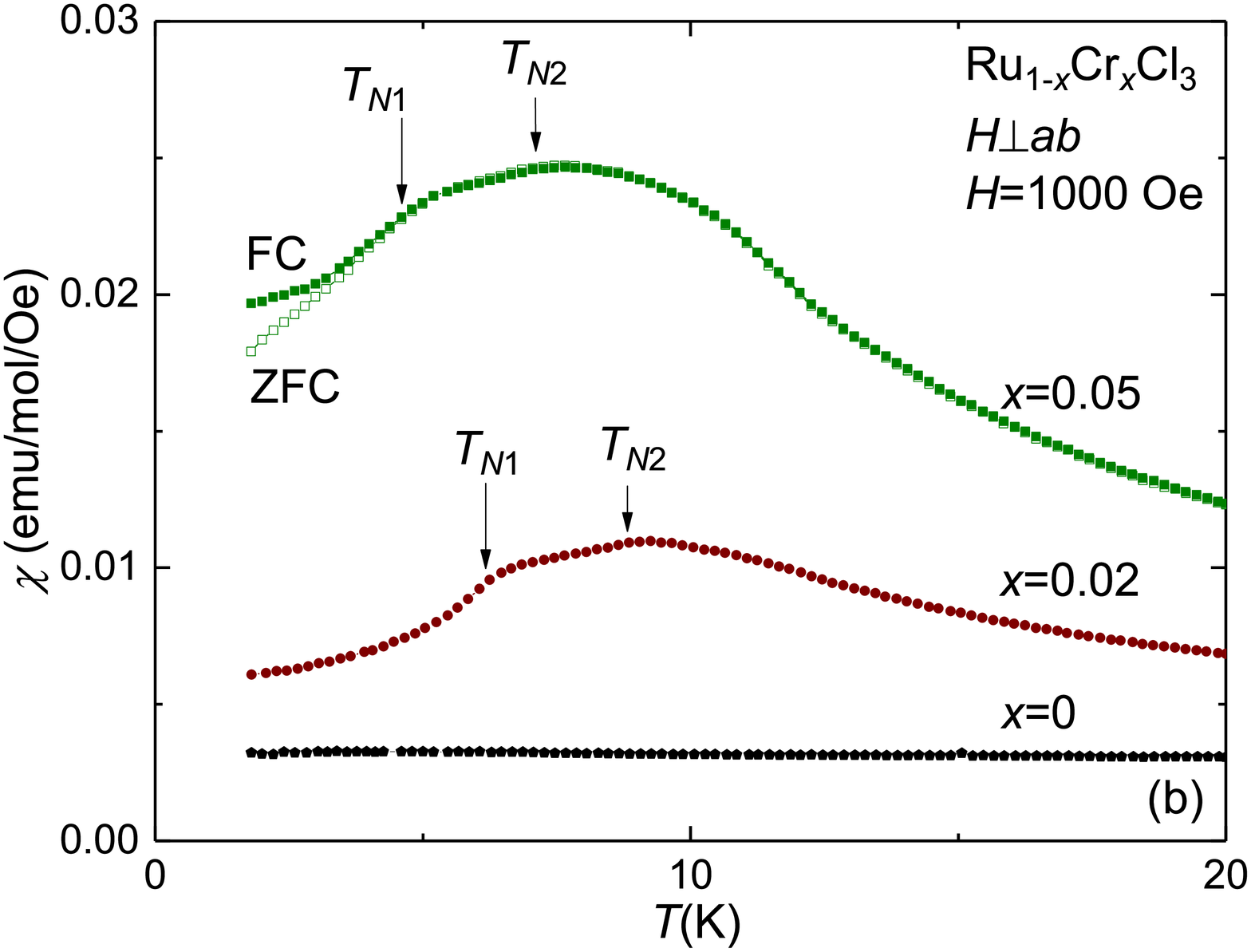}
\caption{(color online)(a) Low temperature Magnetic susceptibility in the $ab$ plane of Ru$_{1-x}$Cr$_x$Cl$_3$ for the substitution ratios $x=0$, $x=0.02$ and $x=0.05$. $T_{N1}$ and $T_{N2}$
indicate the two successive antiferromagnetic transitions and are determined by the maximum of
$d(\chi T)/dT$. (b) Same figure but for a transverse field: $H\bot ab$.} \label{chi_ld}
\end{center}
\end{figure}


In order to have a closer look at the magnetic ordering in the low-doping region, in
Fig.~\ref{chi_ld} the magnetic susceptibility of Ru$_{1-x}$Cr$_x$Cl$_3$ is represented up to 20~K
for small substitution ratios $x = 0$, $x = 0.02$ and $x = 0.05$ both for magnetic fields in and
transverse to the $ab$ plane. Under magnetic fields in the $ab$ plane the magnetization of
$\alpha$-RuCl$_3$ ($x = 0$) shows a sharp transition at $T_{N1}$=7~K indicating the antiferromagnetic
ordering in ABC stacked domains together with a shoulder at $T_{N2}$=10~K hinting at some domains
with stacking faults~\cite{Kubota2015, Cao2016}. At $x = 0.02$ the two magnetic transitions have similar magnitude indicating a strong increase of the stacking disorder under doping in good agreement with previous x-ray diffraction measurements on our single crystals~\cite{Roslova2018}. The magnetic susceptibility for $x = 0.05$ harbors a single broad transition and shows strong similarities with previous measurements on RuCl$_3$
powder~\cite{Kobayashi1992, Johnson2015}, on lower-quality single crystals of
$\alpha$-RuCl$_3$~\cite{Johnson2015} or on Ru$_{1-x}$Ir$_x$Cl$_3$ single
crystals~\cite{Lampen-Kelley2017}. The tiny splitting between zero-field-cooled and field-cooled curves below $T \backsimeq 3$~K might hint at an incipient weak spin-glass behavior at very low temperatures (see section IV).
Thus, the main effect of the substitution of Ru by Cr are an apparent broadening of the
magnetic transition(s), likely due to the induced disorder, and a reinforcement of the second magnetic transition $T_{N2}$ with respect to $T_{N1}$ due to an increased amount of stacking faults in the doped samples. A slight downward shift of the two ordering temperatures $T_{N1}$, $T_{N2}$ indicates the weakening of the zigzag order by the Cr impurities. 

Under a magnetic field transverse to the $ab$ plane, the magnetic susceptibility of
$\alpha$-RuCl$_3$ does not show any clear signature of $T_N$. On the contrary, the magnetic
susceptibility transverse to the $ab$ plane of Ru$_{1-x}$Cr$_x$Cl$_3$ for $x=0.02$ and
$x=0.05$ show clear signatures of the antiferromagnetic transitions under
cooling. Since the magnetic susceptibility transverse to the $ab$ plane appears to be dominated by
the magnetic moments on the Cr site (see discussion above), these measurements indicate the
freezing of the Cr spins in the antiferromagnetic zigzag state for $x=0.02$ and
$x=0.05$.
Under the assumption of an antiferromagnetic Cr-Ru interaction, as
explained above, in the zigzag ordered state the Cr moment would point into the opposite direction
of its two Ru neighbors on the same zigzag structure/leg, with the antiferromagnetic interaction
with the third Ru neighbor not being satisfied. This ordering of Ru and Cr moments is schematically
represented in Fig.~\ref{scheme}(c).





To conclude, the evolution of the magnetic anisotropy in Ru$_{1-x}$Cr$_x$Cl$_3$ as a function of
the Cr content $x$ is remarkable and can be explained by the competition of a strongly anisotropic
Ru-Ru interaction with more isotropic Cr-Cr and Ru-Cr interactions. Overall, our results show, that
the anisotropic $\Gamma$ interactions survive on the Ru-Ru links under partial substitution of Ru
by Cr, i.e., when they are diluted upon Cr substitution via the reduction of the number of Ru-Ru
links in the honeycomb layer.


\subsection{Field dependence of the magnetization}

\begin{figure}[t]
\begin{center}
\includegraphics[width=0.9\linewidth]{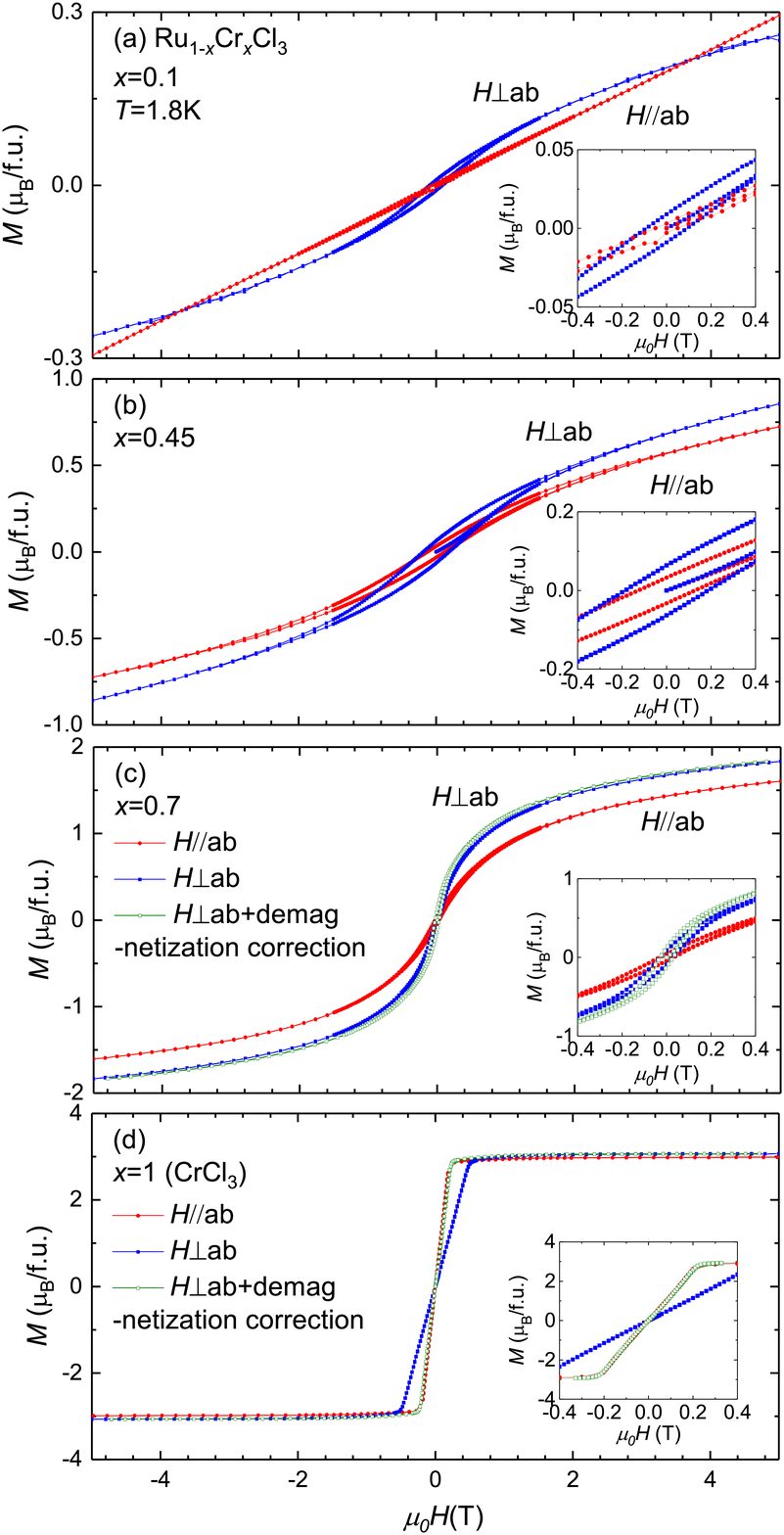}
\caption{(color online)Hysteresis loop of the magnetization of Ru$_{1-x}$Cr$_x$Cl$_3$ at $T$=1.8~K for magnetic field applied in the $ab$
plane and transverse to the $ab$ plane for (a) $x=0.1$, (b) $x=0.45$,
(c) $x=0.7$, (d) $x=1$. The insets magnify the low-field region. The contribution of the demagnetization effect to the transverse
magnetization was evaluated. It is smaller than the symbol size in (a) and (b) and was added to the graph in (c) and (d). For
CrCl$_3$ ($x=1$) the magnetization at $T$=1.8~K after correction of the demagnetization effect does not
show any anisotropy within the error bars. } \label{MvsH}
\end{center}
\end{figure}

The magnetization of Ru$_{1-x}$Cr$_x$Cl$_3$ was measured at 1.8~K as a function of the magnetic
field applied parallel and transverse to the $ab$ plane (see Fig.~\ref{MvsH} for our results for $x=0.1$, $x=0.45$,
 $x=0.7$ and $x=1$). Since the Ru$_{1-x}$Cr$_x$Cl$_3$ single crystals are platelet-like crystals with the a and b axes spanning the plane, the magnetization transverse to the $ab$ plane is reduced by the demagnetization effect. The contribution of the demagnetization effect to the magnetization was estimated for
all compounds assuming an ellipsoidal shape of the sample from Ref.~\cite{Osborn1945}, and was found to
be non negligible for $x=0.7$ and $x=1$. The demagnetization-corrected
magnetization transverse to the $ab$ plane is also plotted in Fig.~\ref{MvsH} for these two compositions.

For $x=0.1$, the magnetization shows a clear deviation from the linear behavior
expected for an antiferromagnetic state with a hysteresis loop up to 1.3~T for both field
directions. It harbors coercive fields of $\mu_0H_c^{ab}$=0.05~T and $\mu_0H_c^\perp$=0.09~T for
magnetic fields parallel and transverse to the $ab$ plane, respectively. While the in-plane
magnetization is close to a linear behavior in field, suggesting strong antiferromagnetic
correlations between the Ru moments, the out-of-plane response shows an S-shape magnetization
curve, probably arising from a small static ferromagnetic component from the Cr moments. For
$x=0.45$, a similar behavior is observed for both field directions, however, with a broadened hysteresis
loop and coercive fields of $\mu_0H_c^{ab}$=0.13~T and $\mu_0H_c^\perp$=0.19~T. For higher substitution
levels, such as for $x=0.7$, the hysteresis gets smaller again with a coercive
field of $\mu_0H_c^{ab}=\mu_0H_c^\perp$=0.03~T. This reduction of the coercive field must come from an
interlayer antiferromagnetic Cr-Cr magnetic interaction in addition to the intralayer ferromagnetic Cr-Cr interaction, similar to the antiferromagnetic interlayer interactions in CrCl$_3$ (see
below). For $x=1$ hysteresis is absent again indicating the overall antiferromagnetic order at
low temperatures in agreement with previous neutron diffraction studies~\cite{Cable1961}. This ordered state is characterized by a ferromagnetic alignment of the spins within
the honeycomb layer and an antiferromagnetic alignment between the layers. After correction of the demagnetization effect, the magnetization of CrCl$_3$ does not show any magnetic anisotropy within our resolution, confirming the absence of an intrinsic magnetic anisotropy in CrCl$_3$ as previously proposed in Ref.~\onlinecite{McGuire2017}.

The full magnetic saturation moment of Ru$_{1-x}$Cr$_x$Cl$_3$ is expected to increase linearly with the Cr content from $M_{sat} \backsimeq 1.3~\mu_B$/f.u. for $\alpha$-RuCl$_3$ for fields parallel to the $ab$ plane~\cite{Johnson2015} to $M_{sat} \approx 3~\mu_B$/f.u for CrCl$_3$ as observed in
Fig.~\ref{MvsH}(d). Notably, the magnetization 5~T of
Ru$_{1-x}$Cr$_x$Cl$_3$ for Cr concentrations $0.1\leq x\leq 0.7$ is still far from the saturation
value. This result indicates that the ferromagnetic Heisenberg interactions between Cr
atoms coexists with antiferromagnetic interactions and hints further at an
antiferromagnetic Ru-Cr interaction within the whole substitution series.

 \begin{figure*}[t]
\begin{minipage}{0.45\linewidth}
\begin{center}
\includegraphics[width=0.98\linewidth]{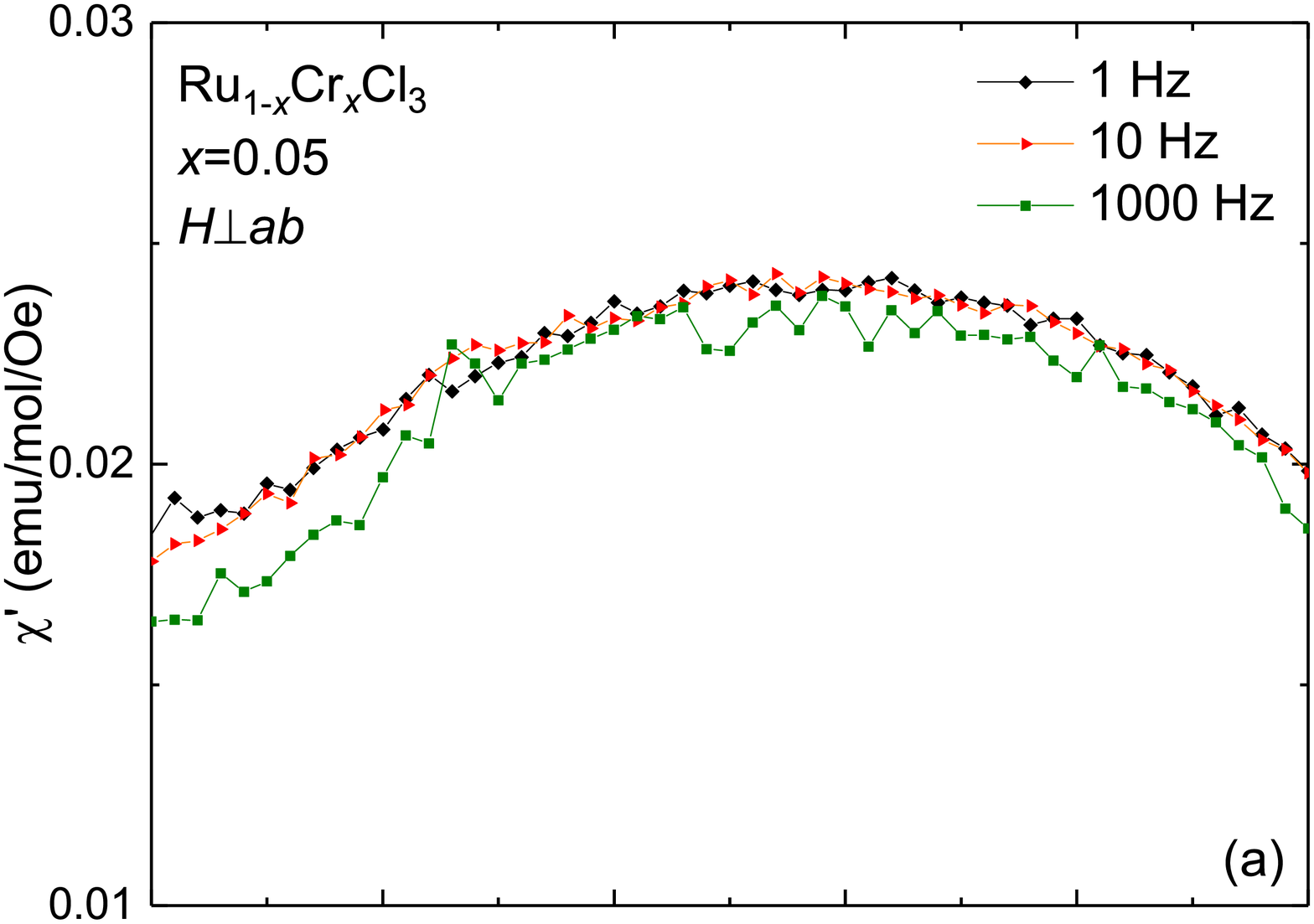}
\includegraphics[width=0.98\linewidth]{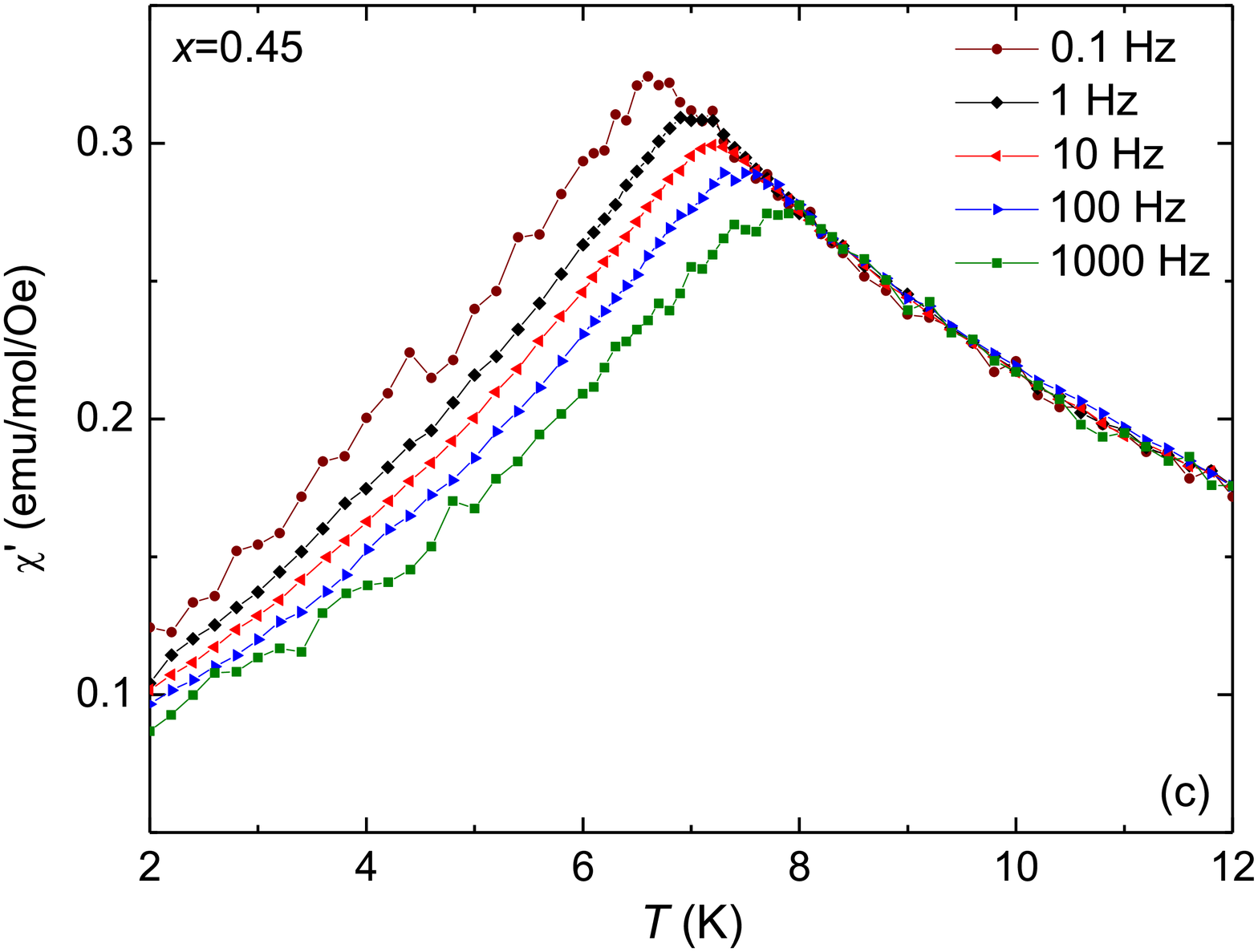}
\end{center}
\end{minipage}
\hfill
\begin{minipage}{0.45\linewidth}
\begin{center}
\includegraphics[width=0.98\linewidth]{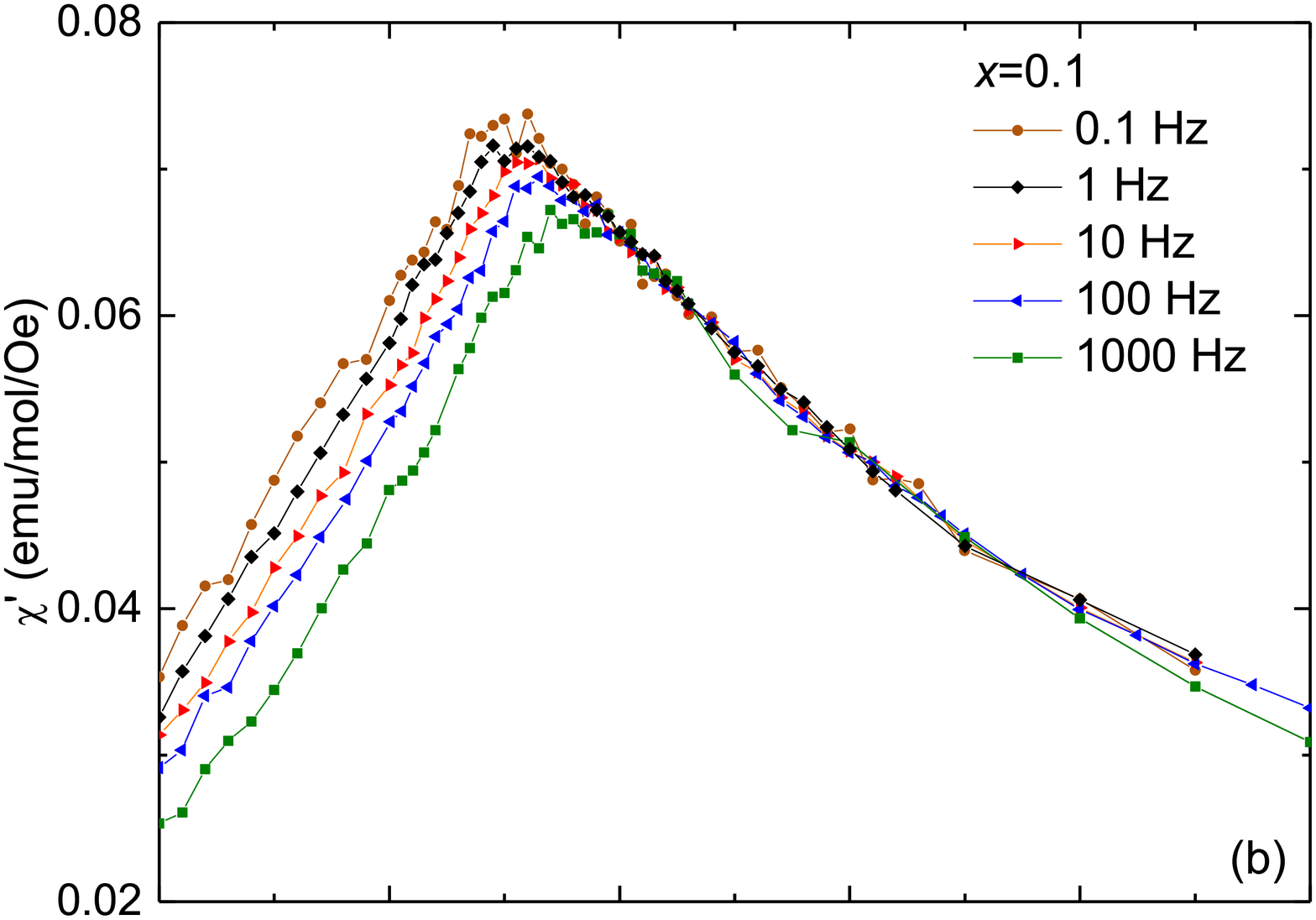}
\includegraphics[width=0.98\linewidth]{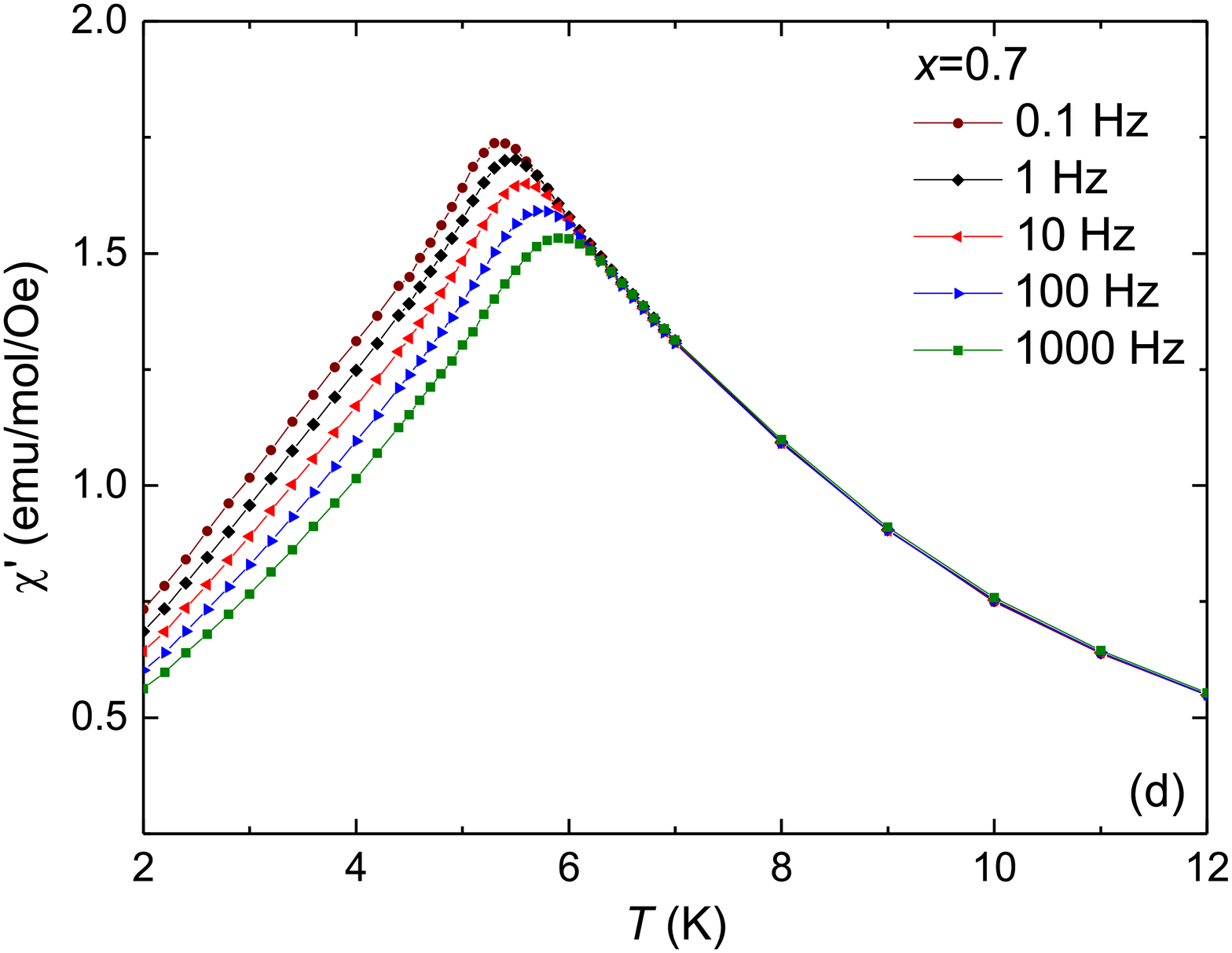}
\end{center}
\end{minipage}
\caption{(Color online) The real part of the ac susceptibility  $\chi'_{ac}$ transverse to the $ab$ plane as a function of
temperature between $T=2~$K and 12~K at different frequencies for Ru$_{1-x}$Cr$_{x}$Cl$_{3}$  $(x =
0.05, 0.1, 0.45, 0.7)$ single crystals. \label{Fig-ac}}
\end{figure*}

\section{AC Magnetic Susceptibility}

The ac susceptibility $\chi_{ac}$ for some of the Ru$_{1-x}$Cr$_{x}$Cl$_{3}$ single crystals was
measured to further elucidate the possibility of a spin-glass state for intermediate substitution
values $x$. The real part of the ac susceptibility $\chi'_{ac}$ is shown in Fig.~\ref{Fig-ac} as a
function of temperature, measured at a fixed $H_{ac} = 5$~Oe transverse to the $ab$ plane, and
$H_{dc} = 0$~Oe, and at various excitation frequencies $f$. For low substitution values ($x =
0.05$), the ac susceptibility shows a broad maximum around $T = 8$~K.
It is frequency independent within the experimental resolution and follows the dc magnetic
susceptibility in Fig.~\ref{chi_ld}(b). This indicates an antiferromagnetic order of Ru and
Cr moments broadened by disorder but no spin-glass state. This is confirmed by the imaginary
component of the ac susceptibility $\chi''_{ac}$, which is zero in the full measured temperature
regime within the experimental resolution (not shown here), as expected for an overall
antiferromagnetic structure
. Note that, the splitting between zero-field-cooled and field-cooled dc magnetization in
Fig.~\ref{chi_ld} for this sample could already hint at an incipient weak spin-glass behavior for $x =
0.05$, not resolved in the ac susceptibility in
Fig.~\ref{Fig-ac}(a).

For the higher substitution levels ($x = 0.1, 0.45, 0.7$) in Ru$_{1-x}$Cr$_{x}$Cl$_{3}$, the ac
susceptibility measurements show clear signatures of a spin-glass transition. A sharp cusp is
observed at a transition temperature (T$_{g}$) at low frequency ($f = 0.1$~Hz), and it matches with
the freezing temperature observed in the dc ($f = 0$) susceptibility measurement. The position of
this cusp is clearly frequency dependent, a classic signature of a spin-glass
transition~\cite{Mydosh1993}. With increasing $x$ the freezing temperature $T_g(x)$
displays a maximum around $x=0.45$. The observation of a spin-glass state up to at least $x=0.7$ indicates, that the magnetic interactions on the remaining Ru-Ru links are still sufficient at $x=0.7$ to induce magnetic frustration.

\section{Heat capacity}

\begin{figure}[t]
\begin{center}
\includegraphics[width=0.9\linewidth]{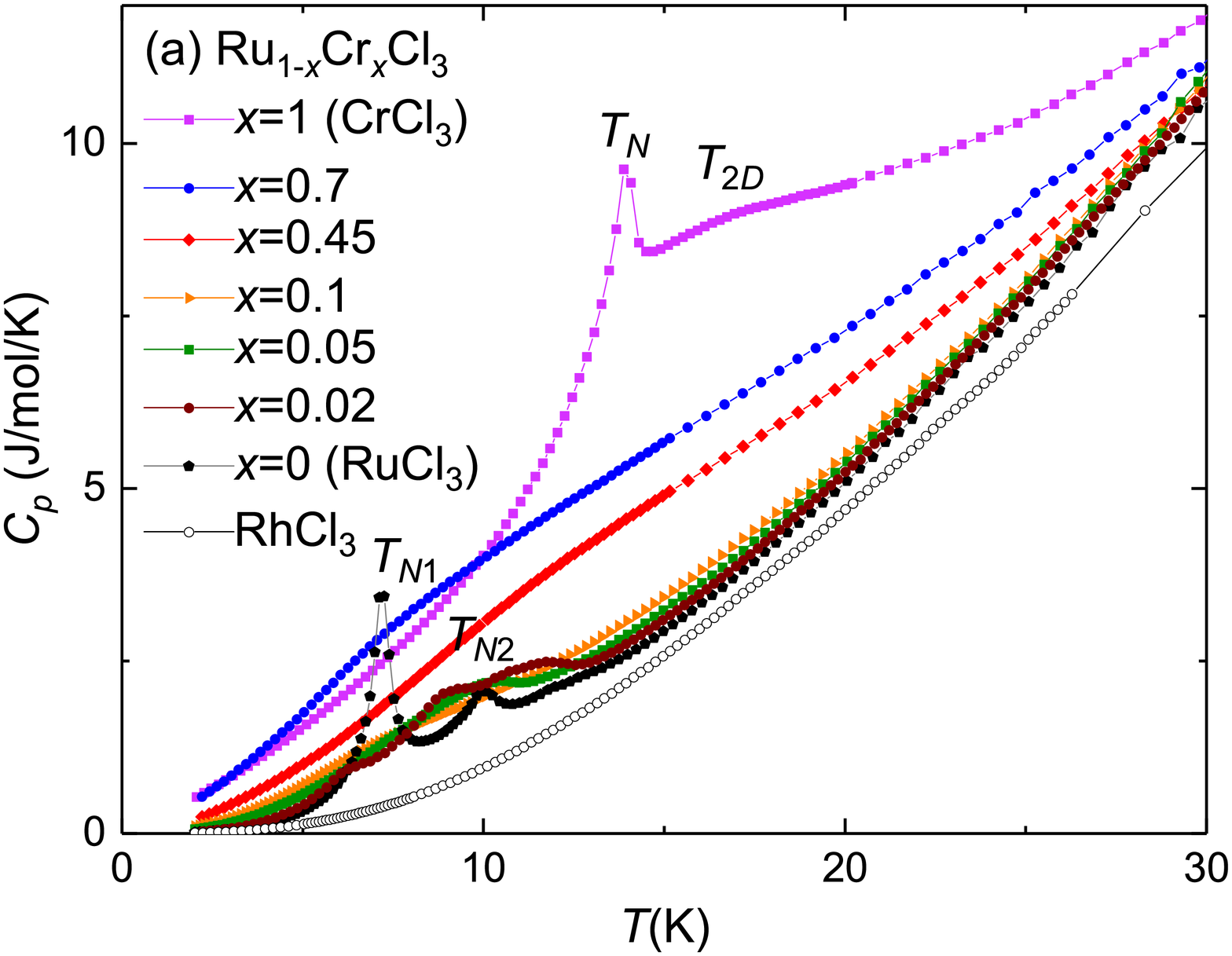}
\includegraphics[width=0.9\linewidth]{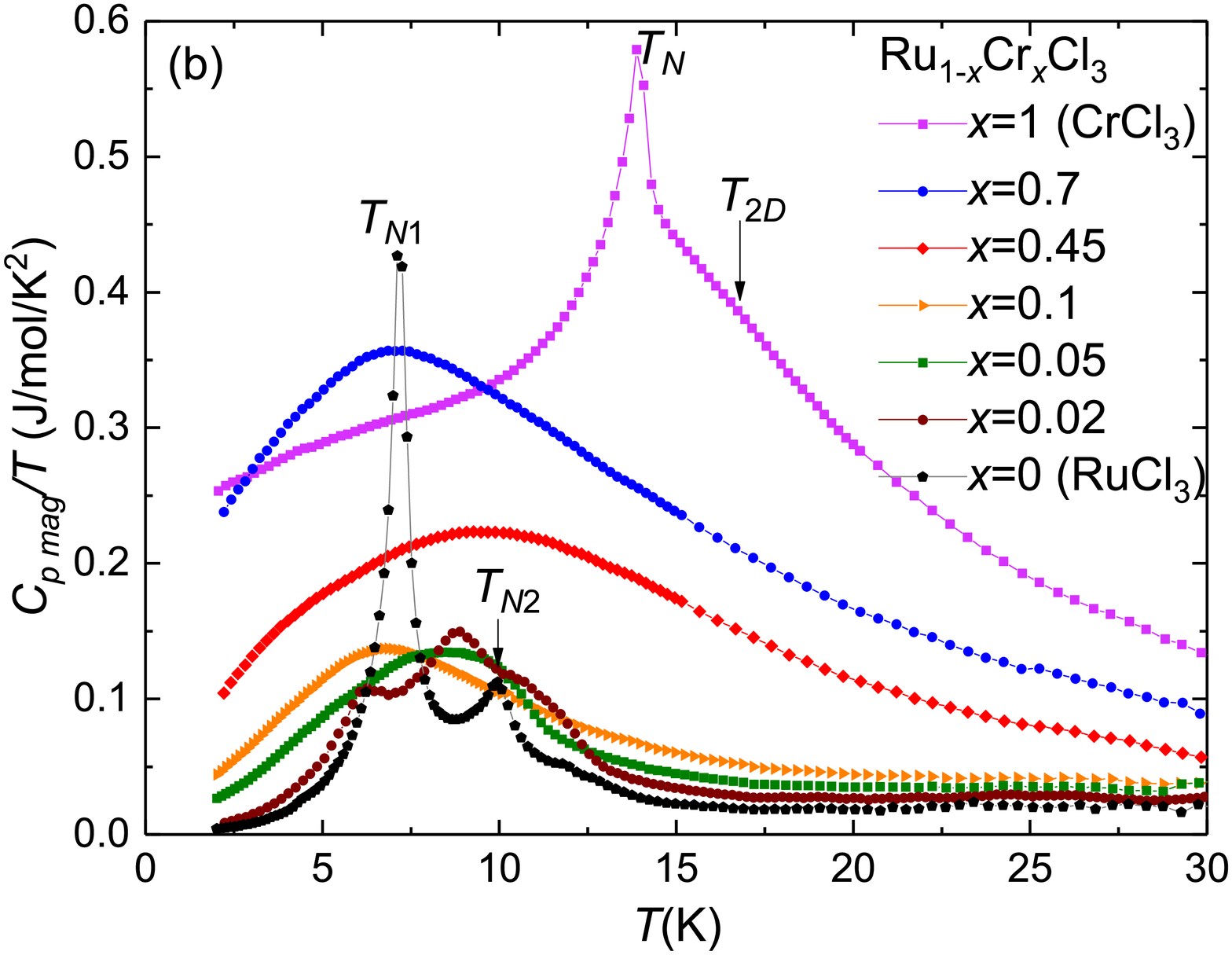}
\caption{(color online)(a) Specific heat of the Ru$_{1-x}$Cr$_x$Cl$_3$ series and the nonmagnetic analog RhCl$_3$
as a function of temperature. (b) Magnetic contribution to the specific heat divided by the
temperature $C_{p,mag}/T$ as a function of temperature. 
} \label{Cp}
\end{center}
\end{figure}

The specific heat of the Ru$_{1-x}$Cr$_x$Cl$_3$ series is represented in Fig.~\ref{Cp} as a
function of temperature together with the specific heat of the nonmagnetic structural analog
compound RhCl$_3$. It was recently suggested from numerical calculations of the phonon spectra of $\alpha$-RuCl$_3$ and RhCl$_3$, that the scaling factor for $\alpha$-RuCl$_3$ should be adjusted to $\theta_D/\theta_{D RhCl3}=0.92$~\cite{Widmann2019}. Such a low scaling factor is in contradiction with our experimental results since it would lead to a negative magnetic contribution to the specific heat in $\alpha$-RuCl$_3$ above $T$=15~K. The phonon contribution to the specific heat for Ru$_{1-x}$Cr$_x$Cl$_3$ was
obtained here by scaling the specific heat of RhCl$_3$ by the Lindemann correction factor taking into
account both the difference in molecular mass and molecular volume between the different
compounds~\cite{Tari2003, Kim2007}. Using the unit cell volume measured by x-ray diffraction for
each sample, which are published in Ref.~\cite{Roslova2018} for Ru$_{1-x}$Cr$_x$Cl$_3$ and $V$=345.19~\AA$^3$ for RhCl$_3$,
 the Lindeman correction factor was estimated to vary with the Cr amount from $\theta_D/\theta_{D RhCl3}=1.00$ for
$\alpha$-RuCl$_3$ to $\theta_D/\theta_{D RhCl3}=1.14$ for CrCl$_3$.



The magnetic contribution to the specific heat, $C_{p,mag}$, was obtained by subtracting the
estimated phononic contribution from the total measured specific heat. The resulting $C_{p,mag}/T$ of Ru$_{1-x}$Cr$_x$Cl$_3$
is represented in Fig.~\ref{Cp}(b) as a function of temperature. For $x =
0$ and
$x =
0.02$, the specific heat shows the same two transitions $T_{N1}$ and
$T_{N2}$ as the magnetization, corresponding to two different stacking sequences, again confirming
the predominance of $T_{N1}$ for $x =0$ and of $T_{N2}$ for
$x =0.02$. Note that the two magnetic transitions are broadened and shifted to
lower temperature for Cr-substituted $x =0.02$. For $x =0.05$ we
finally measured a single broad transition. Like for the magnetization, strong similarities can be
observed between the specific heat for $x =0.05$ and previous measurements on
RuCl$_3$ powder~\cite{Banerjee2016}, on low-quality single crystals of
$\alpha$-RuCl$_3$~\cite{Cao2016} or on Ir-doped single crystals~\cite{Do2018}. Thus, this very
broad transition for $x =0.05$ indicates an enhanced disorder inside the layer and in the layer stacking
upon doping and might lead to a change from a long-range antiferromagnetic
zigzag order in $\alpha$-RuCl$_3$ to a short-range order in the doped samples. This phenomenon has
also been observed in the Ir-substitution series Ru$_{1-x}$Ir$_x$Cl$_3$~\cite{Do2018} via a
combined specific heat and muon spectroscopy study. This short-range order might already harbor a
spin-glass behavior, despite the fact that it could not be clearly resolved in our ac
susceptibility measurements.

For higher Cr concentrations, $0.1\leq x\leq0.7$, where the dc and ac magnetic susceptibility clearly indicate a spin-glass state, specific-heat measurements confirm the
absence of long-range magnetic order. $C_{p,mag}/T$ as a function of temperature shows a broad bump with a maximum around $T=7-9$~K i.e. slightly above the spin-glass freezing temperature $T_g$. This bump indicates an entropy loss via the formation of short-range correlation. Its maximum is at slightly higher temperature for $x = 0.45$, than for lower Cr content ($x = 0.1$) or higher Cr content ($x = 0.7)$.
An overall increase of the magnetic contribution to the specific heat is observed as function of
the Cr content $x$ for $0.05 \leq x \leq 1$, which is expected for the substitution of $J_{\rm eff} =
1/2$ by $S = 3/2$ spins.

For $x=1$ (CrCl$_3$) our specific-heat study further confirms the antiferromagnetic transition at $T_N =
14.1$~K. In addition, a broad bump around $T_{2D} = 17$~K may indicate the 2D ferromagnetic
ordering of CrCl$_3$ proposed in Ref.~\onlinecite{Kuhlow1982}, although for our crystals this transition is much less pronounced than in a previous specific study on CrCl$_3$~\cite{McGuire2017}.
Interestingly, our measurements show a significant magnetic contribution to the specific heat in
CrCl$_3$ down to 2~K. This low-$T$ magnetic entropy far below the N\'eel temperature $T_N = 14.1$~K is likely due to strong fluctuations in this quasi-two-dimensional magnet.
At first glance, our results disagree with a previous specific-heat study of CrCl$_3$~\cite{McGuire2017}, but where the
phononic contribution to the specific heat was estimated assuming $C_{p,mag}/T$ = 0 for $T <
7.5$~K, i.e., a different analysis has been applied. The present study with the use of a
nonmagnetic structural analog compound allows to compute the magnetic entropy of CrCl$_3$ with a
higher accuracy, yielding a magnetic entropy $S_{mag}(30~K)-S_{mag}(2~K)=\int_2^{30}{C_{p,mag}/T}
dT = 8.5$~J/mol/K, which is only slightly smaller than the expected value for a $S = 3/2$ system
($S_{mag,theo} = R\ln4$ = 11.5~J/mol/K).



\section{x-T phase diagram}

\begin{figure}[t]
\begin{center}
\includegraphics[width=0.9\linewidth]{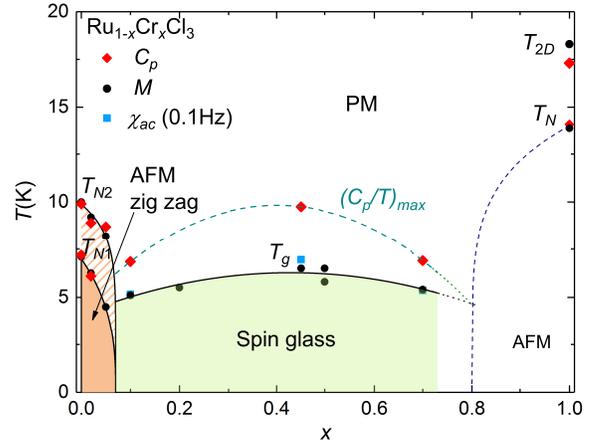}
\caption{(color online)$(x-T)$ phase diagram of Ru$_{1-x}$Cr$_x$Cl$_3$. The black circles, red diamonds and blue
squares are experimental points from magnetization, specific-heat and ac magnetic susceptibility
($f = 0.1~$Hz) measurements, respectively. The lines are guides to the eye. The orange and the
orange and white patterned region corresponds to the fully ordered zigzag state and the region of
the phase diagram, where only the part of the sample with numerous stacking faults is magnetically
ordered, respectively. The dashed green line indicates the temperature where $dS_{mag}/dT =
C_{p,mag}/T$ undergoes its maximum. The boundary between the spin-glass state and the
antiferromagnetic state on the high Cr content side could not precisely be determined in this study and is thus schematically drawn, only.}
\label{xT}
\end{center}
\end{figure}

The experimental results reported in the three previous sections enable us to draw the ($x-T$)
phase diagram of Ru$_{1-x}$Cr$_x$Cl$_3$ (Fig.~\ref{xT}). The two N\'eel temperatures $T_{N1}$ and
$T_{N2}$ decrease with the Cr content $x$ showing that the Cr $S$ = 3/2 impurities disfavor the
antiferromagnetic zigzag order of $\alpha$-RuCl$_3$, finally leading to a destabilization of the
zigzag order in favor of a spin-glass state for a low Cr content $x \backsimeq
 0.1$. This spin-glass state is stable for a broad Cr doping interval up to at
least $x = 0.7$, with a maximum of the freezing temperature $T_g$ around $x \backsimeq 0.45$.
Due to the proximity of $\alpha$-RuCl$_3$ to the Kitaev spin liquid, this spin glass state may harbor bound states of Majorana Fermions as low-energy spin excitations.
The transition from the
spin-glass state toward the antiferromagnetic order of CrCl$_3$ in the high Cr concentration limit
could not be investigated in detail in this study due to a lack of suitable crystals. The freezing temperature may evolve continuously toward a second-order magnetic transition upon Cr substitution from $x = 0.7$ to $x = 1$. A minimum of the freezing/ordering temperature must occur between $x = 0.7$ and $x = 1$, and the origin and nature of this minimum need to be elucidated in detail in future studies.

\section{Summary and conclusion}

We have studied the magnetic properties of the substitution series Ru$_{1-x}$Cr$_x$Cl$_3$ by
magnetization, ac susceptibility and specific-heat measurements. In the low-substitution limit $x
\leq 0.05$ the moments on the Cr sites can be considered as magnetic impurities in a Kitaev
magnet. These impurities are antiferromagnetically coupled to the neighboring Ru
moments. In an out-of-plane magnetic field, where the response of the Ru moments is small due to the off-diagonal $\Gamma$ interactions, the Cr impurity moments can easily be tilted because of the small mean field resulting from the frustrated parent state. The unusual evolution of the magnetic anisotropy, showing a reversal of the anisotropy for relatively small $x$, can be successfully explained via the competition between isotropic Heisenberg and anisotropic magnetic
$\Gamma$ interactions 
in this series of frustrated magnets.
At low temperatures, the substitution of Ru by Cr results in an evolution of the antiferromagnetic
zigzag order into a possible short-range ordered state around $x \backsimeq 0.05$, and then into a
spin glass for $x \backsimeq 0.1$ indicating that the antiferromagnetic long-range order of
$\alpha$-RuCl$_3$ is not robust under the substitution of Ru atoms by magnetic Cr impurities.
The evolution of the magnetic properties for higher doping levels shows the gradual development from a system dominated by the Kitaev $K$ and off-diagonal $\Gamma$ interactions toward a system dominated by ferromagnetic Cr-Cr interactions of Heisenberg-type. However, anisotropic 
$\Gamma$ interactions on Ru-Ru nearest-neighbor links appear to survive in the whole substitution series. The spin-glass state is observed for a broad Cr content interval up to at least $x \backsimeq  0.7$ with a maximum of the freezing temperature around $x = 0.45$.

It should be mentioned that the scenario of isotropic (Heisenberg) Ru-Cr interactions represents a minimal model to explain our experimental observations for the magnetic anisotropy along the series Ru$_{1-x}$Cr$_x$Cl$_3$. A detailed numerical modelling of susceptibilities is required to resolve possible anisotropies of the Ru-Cr interactions; such studies are currently under way.
In addition, a careful characterization of samples at very small doping is called for, in particular also at elevated fields, in order to connect to theoretical investigations of single-impurity effects in the quantum limit and to probe the stability of the proposed field-induced QSL state.

\begin{acknowledgments}
Insightful discussions with E. Andrade, L. Hozoi, L. Janssen, V. Kataev, and S. Nagler as well as technical support by S. Gass is acknowledged.
We acknowledge financial support from the DFG through SFB 1143 (project-id 247310070) and the W\"urzburg-Dresden Cluster of Excellence on Complexity and Topology in Quantum Matter -- \textit{ct.qmat} (EXC 2147, project-id 39085490) as well as from the
European Union's Horizon 2020 research and innovation programme under the Marie Sk\l{}odowska-Curie
grant agreement No 796048. K. Mehlawat acknowledges the Hallwachs-R\"ontgen Postdoc Program of \textit{ct.qmat} for financial support.
\end{acknowledgments}

%

\end{document}